\documentclass[twocolumn,showpacs,nofootinbib,tightenlines,nobibnotes, aps, amssymb,amsmath,prb]{revtex4-1}
\usepackage{graphicx}
\usepackage{epsfig}
\usepackage{amssymb,amsmath,amsfonts,hyperref}
\usepackage{wasysym}
\usepackage{latexsym}
\usepackage{eucal}
\usepackage{verbatim}
\usepackage[normalem]{ulem}
\usepackage{enumitem}
\setlist{nosep}

\usepackage{color}

\newcommand{\tetr}{\Box}
\begin{document}

\title{Efficient quantum cluster algorithms for frustrated transverse field Ising antiferromagnets and Ising gauge theories.}
\author{Sounak Biswas}
\affiliation{\small{Tata Institute of Fundamental Research, 1 Homi Bhabha Road, Mumbai 400005, India}}
\author{Kedar Damle}
\affiliation{\small{Tata Institute of Fundamental Research, 1 Homi Bhabha Road, Mumbai 400005, India}}
\begin{abstract}
  Working within the Stochastic Series Expansion (SSE) framework, we construct efficient quantum cluster algorithms for transverse field Ising antiferromagnets on the pyrochlore lattice and the planar pyrochlore lattice,  for the fully frustrated square lattice Ising model in a transverse field (dual to the 2+1 dimensional odd Ising gauge theory), and for a transverse field Ising model with multi-spin interactions on the square lattice, which is dual to a 2+1 dimensional even Ising gauge theory (and reduces to the two dimensional quantum loop model in a certain limit). Our cluster algorithms use a microcanonical update procedure that generalizes and exploits the notion of ``pre-marked motifs'' introduced earlier in the context of a quantum cluster algorithm for triangular lattice transverse field Ising antiferromagnets. We demonstrate that the resulting algorithms are significantly more efficient than the standard link percolation based quantum cluster approach. We also introduce a new
canonical update scheme that leads to a further improvement in measurement of some observables arising from its ability to make one-dimensional clusters in the ``imaginary time'' direction. Finally, we demonstrate that refinements in the choice of premarking strategies can lead to additional improvements in the efficiency of the microcanonical updates. As a first example of the physics that can be studied using these algorithmic developments, we obtain evidence for a power-law ordered intermediate-temperature phase associated with the two-step melting of long-range order in the fully frustrated square lattice transverse
field Ising model.

\end{abstract}

\vskip2pc

\maketitle
\section{Introduction}
Transverse field Ising antiferromagnets on frustrated lattices (and closely
related quantum dimer, quantum link models and anisotropic spin models) have served as paradigmatic
examples of the interplay between quantum fluctuations, thermal fluctuations, and the entropic effects associated with a large
degeneracy of configurations with low exchange energy.\cite{Moessner_Sondhi_PRB,Moessner_Sondhi_Chandra_PRL,Nikolic_Senthil,Nikolic,Hermele_Balents_Fisher,Banerjee_etal_PRL,Roechner_Balents,Moessner_Sondhi_RVBQDMIGT_review,Moessner_Sondhi_Z2,Ralko_Z2,Isakov_Moessner,Mila_fftfim,Moessner_Sondhi_Chandra_QDM,Schlittler_etal_QDM,Syljuasen_Chakravarty,Shannon_Misguich_Penc,Henry_Roscilde,Banerjee_etal_JStat,Banerjee_PRB,Banerjee_PoS}
They have been used to study a host of interesting
phenomena associated with this interplay. Examples include Z$_2$ spin liquid phases realized in a variety of two dimensional
settings,\cite{Moessner_Sondhi_Z2,Ralko_Z2} a U(1) Coulomb liquid phase realized on the three dimensional
pyrochlore lattice,\cite{Hermele_Balents_Fisher,Banerjee_etal_PRL,Roechner_Balents} and
 long range
order established at low temperature via order-by-disorder effects.\cite{Isakov_Moessner,Mila_fftfim,Moessner_Sondhi_Chandra_QDM,Schlittler_etal_QDM,Syljuasen_Chakravarty,Shannon_Misguich_Penc,Henry_Roscilde,Banerjee_etal_JStat,Banerjee_PRB,Banerjee_PoS}

Although ground state methods such as diffusion Monte Carlo and exact diagonalization have been used
to study the quantum phases of of such model Hamiltonians and related quantum dimer and link models,\cite{Ralko_Z2,Syljuasen_Chakravarty,Shannon_Misguich_Penc} efficient cluster algorithms for completely unbiased Monte Carlo studies of the $T \geq 0$ quantum statistical mechanics of these systems have been largely unavailable in spite of the important role played
by such models in the theory of frustrated quantum systems. Continuous-time cluster algorithms have been formulated for transverse field Ising models previously.\cite{Sandvik_SSE,Rieger_Kawashima} However, these methods typically construct clusters based on a ``\emph{link-decomposition}'' of the Hamiltonian into terms living on the links of the spatial lattice. While these cluster algorithms work well for ferromagnetic systems, they are rather inadequate for frustrated systems, since clusters formed by them percolate and freeze. Studies of models with frustration have mostly relied on growing one-dimensional clusters in the imaginary time direction.\cite{Isakov_Moessner,Mila_fftfim,Nakamura} Recently, cluster algorithms have been formulated for three dimensional
classical statistical mechanics problems equivalent in the $\tau$-continuum limit to some specific transverse field Ising models,\cite{Henry_Roscilde} or quantum dimer models\cite{Banerjee_etal_JStat, Banerjee_PRB, Banerjee_PoS}. These require an extrapolation to the $\tau$-continuum limit to accurately compute the physics of the quantum system.

A recent exception is a quantum cluster algorithm\cite{Biswas_Rakala_Damle}  developed 
for a class of transverse field Ising antiferromagnets on the triangular lattice with
subdominant further neighbour couplings of either sign. This algorithm uses a plaquette decomposition of the Hamiltonian within the Stochastic
Series Expansion (SSE) framework of Sandvik\cite{Sandvik_SSE,Sandvik_TFIM} to construct clusters based on a plaquette percolation process. Flipping these clusters
does not change the weight of a configuration. For triangular antiferromagnets, this microcanonical cluster update is significantly more efficient than the standard link percolation based microcanonical cluster construction developed in
Ref.~\onlinecite{Sandvik_TFIM}. This enhanced performance has been recently exploited\cite{Biswas_Damle_thermodynamic} to
study a thermodynamic signature of two-step melting\cite{Damle_melting_PRL} of three-sublattice order.

In the case of bosonic systems with frustrated interactions (and equivalent XXZ spin models), merely working with an appropriate cluster decomposition of the Hamiltonian suffices\cite{Louis_Gros,Kedar_Dariush}
to dramatically improve the efficiency of directed loop updates\cite{Sandvik_loopupdate,Syljuasen_Sandvik} within the SSE framework.
As was already noted in Ref.~\onlinecite{Biswas_Rakala_Damle}, this does not suffice for SSE simulations of frustrated transverse field Ising models. The other key ingredient needed in such systems which have no conserved charge is the notion of {\em premarked motifs} that are imprinted on each spatial plaquette at the start of the cluster construction
process. In the algorithm of Ref~\onlinecite{Biswas_Rakala_Damle}, these premarked motifs determine
the manner in which legs of {\em all} diagonal plaquette
operators ``living on'' a given spatial plaquette get assigned to ``space-time clusters''. Once all clusters are constructed, each can be independently flipped with probability half within a Swendsen-Wang type\cite{Swendsen_Wang} implementation. This premarking based cluster construction leads to a certain consistency in the way operators are split into different clusters all along the SSE operator string. This consistency leads
to vastly improved cluster size statistics. The resulting broad distribution of cluster
sizes is the real key to the efficiency of the quantum cluster algorithm developed in Ref.~\onlinecite{Biswas_Rakala_Damle}.
This microcanonical cluster algorithm\cite{Biswas_Rakala_Damle} reduces, in the limit of
vanishing transverse field, to a variant of the Kandel-Ben Av-Domany cluster
construction for the corresponding classical Ising model,\cite{KBD_PRL,Kandel_Domany,KBD_PRB,Coddington_Han,Zhang_Yang}  which constructs
clusters starting with a decomposition of the classical exchange energy into
terms that live on triangular plaquettes. The notion of premarking however leads to a key difference from the standard Kandel-Ben Av-Domany  construction, and provides an efficient way to generalize these classical ideas to the quantum model.

In this paper, we exploit this insight to develop microcanonical cluster updates within the SSE framework for
transverse field Ising antiferromagnets on the pyrochlore lattice and the planar pyrochlore lattice,  the fully frustrated square lattice Ising model  in a transverse field (dual to the 2+1 dimensional odd Ising gauge theory), and a transverse field Ising model (TFIM) with multi-spin interactions on the square lattice, which is dual to a 2+1 dimensional even gauge theory and reduces to the two dimensional quantum loop model in a certain limit. 

We demonstrate that these quantum cluster algorithms are significantly more efficient than the standard link percolation based quantum cluster approach of Ref.~\onlinecite{Sandvik_TFIM}. We also introduce a new
canonical update scheme and show that it leads to a further improvement in autocorrelation times for some observables. 
Finally, we demonstrate that refinements in the choice of pre-marked motifs can lead to additional and very significant improvements in efficiency in certain cases.
These refinements are of two types. First, one can use some intuition to try and choose the pre-marked motifs to be ``compatible'' with the underlying equilibrium ensemble.  Second, and somewhat
more involved in terms of implementation, is to choose the pre-marked motifs to be
``compatible'' with the current system configuration in a manner that
preserves detailed balance. This second approach also involves switching to a Wolff-type\cite{Wolff}
variant of the cluster algorithm, in which a single cluster is grown from a random starting
point and then flipped with probability one. As a first example
of the physics that can be studied using these algorithmic developments, we obtain evidence for a power-law ordered intermediate temperature phase associated
with the two step melting of long-range order in the fully frustrated square lattice transverse
field Ising model.

The rest of this paper is organized as follows: In Sec.~\ref{models}, we define the
four different transverse field Ising models which we study here, and provide a heuristic description of the low temperature physics expected in each of these models. In Sec.~\ref{review}, we
provide a brief introduction to the SSE framework, with particular emphasis on the standard link percolation based cluster algorithm of Ref.~\onlinecite{Sandvik_TFIM}. In Sec.~\ref{microcanonical}, we introduce our prescription for the choice of motifs (and the cluster decomposition rules implied by these motifs) for our plaquette percolation based construction of microcanonical clusters in each of these four models. We also describe various premarking strategies to imprint these motifs on the underlying spatial lattice.  In Sec.~\ref{canonical}, we introduce a {\em canonical}  update procedure which leads to further improvements in Monte Carlo autocorrelation times for some observables. This is followed by a discussion of the performance
of our methods in various cases in Sec.~\ref{Performance}. Finally, we illustrate the power
of our methods in Sec.~\ref{Results} by reporting results
that provide strong evidence for a power-law ordered intermediate phase
associated with the two-step melting of the ground state columnar
order in the small transverse field regime of the fully frustrated transverse
field Ising model. 

\begin{figure}[t]
  \includegraphics[width=6cm]{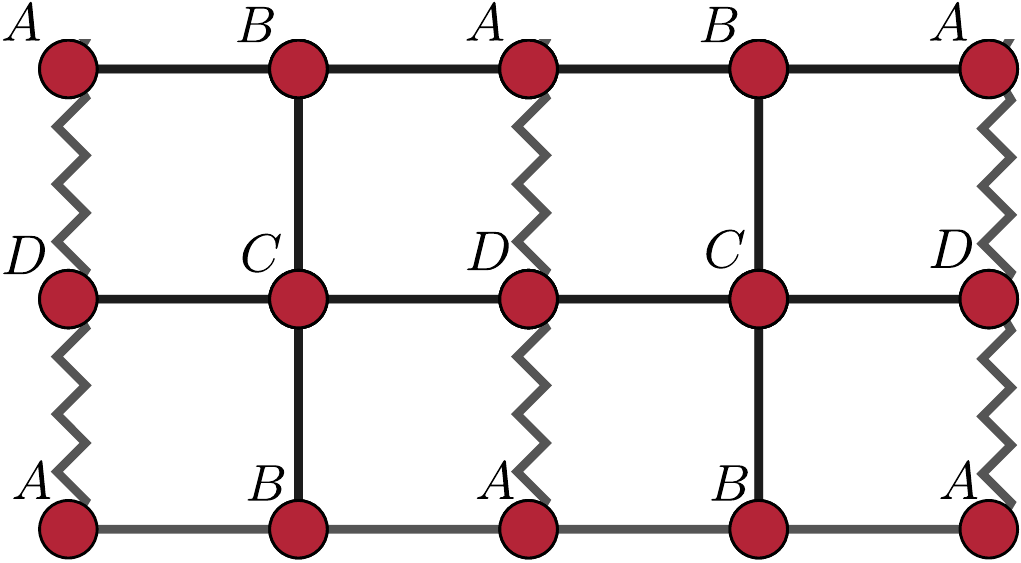}
  \caption{\label{fftfim_lat}The fully frustrated Ising model on the square lattice. The straight lines denote ferromagnetic couplings, while the zigzag lines denote antiferromagnetic couplings. The choice of antiferromagnetic bonds is not unique, and constitutes a gauge choice. The site labels denote the four sublattice structure of the square lattice..}
\end{figure}
\begin{figure}[t]
\includegraphics[width=\columnwidth]{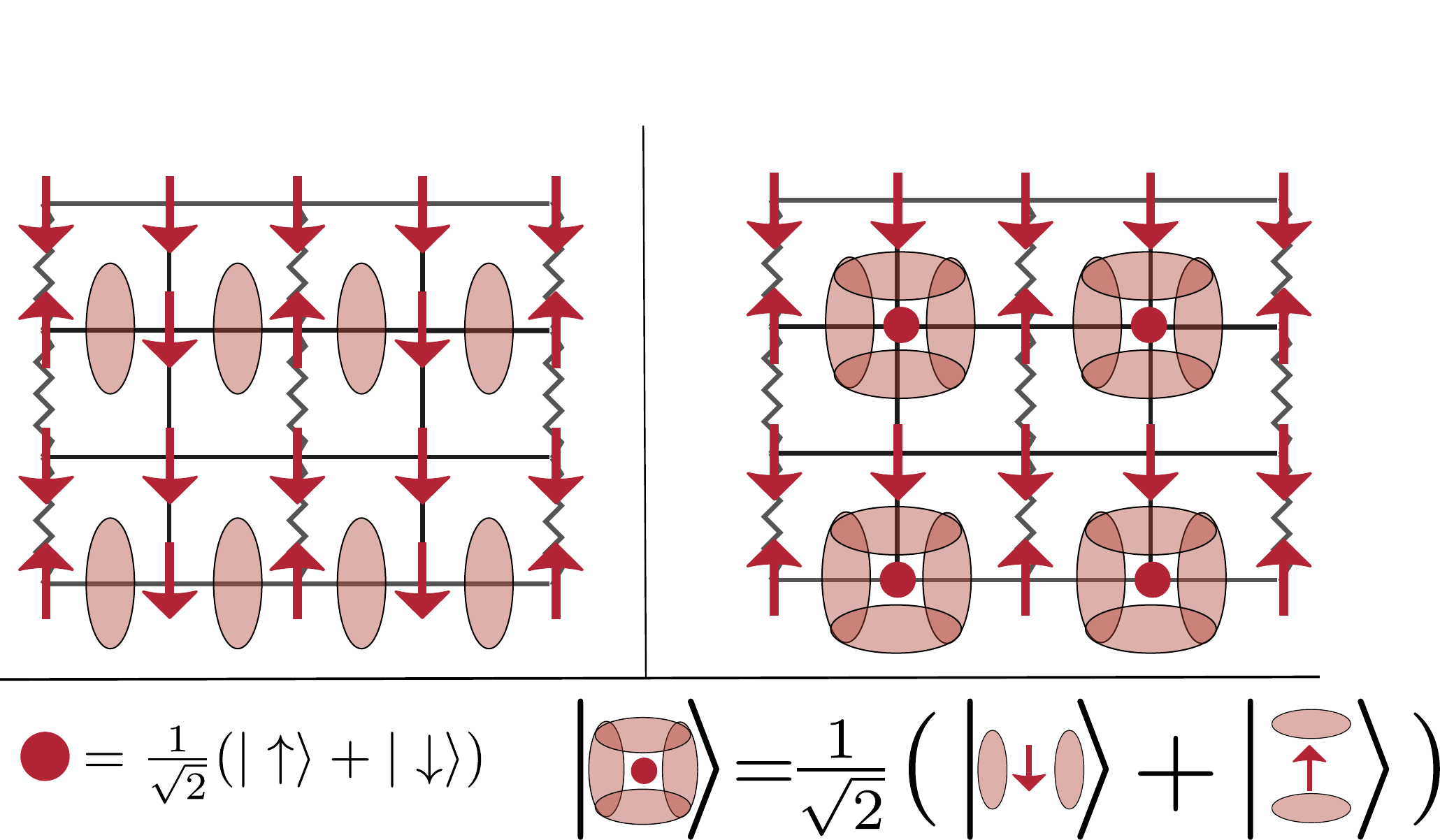}
\caption{\label{OFIM_cartoon} Schematic representation of columnar ordered
(left) and plaquette ordered (right) ground states of the fully frustrated Ising model in a transverse field. There are four symmetry related columnar states and four symmetry related plaquette ordered states, of which one representative each has been shown in figure.} 
\end{figure}
\section{Models}
\label{models}
\subsection{Fully frustrated TFIM on the square lattice.}
The fully frustrated TFIM on the square lattice is given by the Hamiltonian
\begin{equation}
  H_{\mathrm TFIM}= -\sum_{<ij>}J_{ij}\sigma^z_i \sigma^z_j -\Gamma \sum_{i}\sigma^z_i.
  \label{fftfim_eqn}
\end{equation}
Here $i$ labels the sites and $\langle ij\rangle$ labels the nearest neighbors links of the square lattice.
The nearest neighbour Ising exchange interactions $J_{ij}$ has been shown in Fig.~\ref{fftfim_lat},
with $J_{ij}=|J|$ on the straight lines, and $J_{ij}=-|J|$ on the zig-zag lines. Thus, the product of the exchange couplings around an elementary plaquette is always negative, {\em i.e.}
each plaquette is threaded by $\pi$ flux'. While this is a gauge invariant characterization
of the fully-frustrated nature of the interactions, the actual choice of $J_{ij}$ is not
unique, and corresponds to a gauge freedom. Spin
correlations naturally depend on this gauge choice, while bond energy correlations
are gauge-invariant, {\em i.e.} independent of the way in which the fully frustrated interaction
is implemented. Due to the fully frustrated nature of the interactions, there is a macroscopic
entropy of spin configurations that minimize the Ising exchange energy. In each such
configuration, each plaquette of the square lattice hosts exactly one frustrated bond (on which the exchange interaction energy is not minimized).
Higher energy configurations also have plaquettes with three frustrated bonds. 

The bond energies can be used to define dimer variables on links of the dual square lattice, with a frustrated bond  represented by a dimer on the dual link. In the dimer language, all dual sites have either one or three dimers touching them. Assigning
an Ising variable $\tau^x=-1$ to each dual link with a dimer, and $\tau^x=+1$ to each empty dual link, we see that this is precisely the configuration space of the odd Ising
gauge theory, in which the Gauss law constraint requires that the product
of $\tau^x$ on all links meeting at a dual site be $-1$. The transverse field
term enforces a cost of $2\Gamma$ for a vison on any dual plaquette (a vison corresponds to the product of $\tau^z$ around the plaquette being $-1$),
while the Ising exchange term translates to a uniform transverse field of strength $|J|$ coupling to $\tau^x$.

Minimum exchange energy
configurations correspond to perfect dimer covers (with each dual site touched by
exactly one dimer) of the dual lattice. In the limit of small $\Gamma/|J|$, such
configurations are expected to dominate. To lowest order in degenerate perturbation
theory in $\Gamma/|J|$, the transverse field term corresponds to a ring exchange
term for the dimers, flipping a pair of horizontal dimers on a dual plaquette to pair of
vertical dimers. This is just the quantum dimer model with the usual ring exchange term. In this language, higher orders in this perturbation expansion are expected to generate additional kinetic terms, as well as interactions between dimers.

For small $\Gamma$, this suggests two natural possibilities for the ground state: The first is
a columnar ordered state, in which dimers line up along columns and break lattice translation symmetry either along the $x$ axis or the $y$ axis, but not both. The second is
a plaquette ordered state, in which pairs of dimers resonate between two orientations
on one quarter of the dual plaquettes, breaking symmetry both along the $x$
and the $y$ directions. Cartoons corresponding to these possibilities are displayed in Fig.~\ref{OFIM_cartoon}. The latter of these two possibilities is analogous to the antiferromagnetic three-sublattice ordered ground state that is realized for small transverse fields in the transverse field Ising antiferromagnet on the triangular
lattice\cite{Isakov_Moessner,Biswas_Rakala_Damle,Biswas_Damle_thermodynamic} while the former of the two is the analog of the density wave ordering seen in the $S=1/2$ triangular XXZ model\cite{Kedar_Dariush,Wessel_Troyer,Bonninsegni_Prokofiev} with antiferromagnetic exchange couplings for the $z$ components and ferromagnetic transverse exchange couplings. It is also the ordering pattern stabilized\cite{DPLandau} in the classical Ising antiferromagnet by a small second neighbour ferromagnetic interaction between Ising spins. 

Recent quantum Monte Carlo computations\cite{Mila_fftfim} have provided strong evidence in favour of a columnar ordered state at low values of $\Gamma$, separated from
the high $\Gamma$ quantum paramagnet by a single quantum phase transition at $T=0$.
Given this ordered state, one expects (on Landau theory grounds) two transitions as a function of increasing temperature in the small $\Gamma$ regime. The first is from the ordered state to a state with quasi-long range order (power-law spin correlations) and continuously varying power-law exponents (corresponding to a fixed line in renormalization group language), while the second is from this power-law phase to a high temperature paramagnet. However, previous computational work\cite{Mila_fftfim} has not been able to provide any direct confirmation of this interesting two-step melting scenario.
Another interesting aspect of this small $\Gamma$ physics is the anomalously large
crossover length scale below which the system apparently fails to distinguish between
plaquette and columnar ordered possibilities even far away from the quantum
phase transition to the large $\Gamma$ paramagnet. This crossover length scale gets even larger closer to the quantum phase transition, making a computational
study of this phenomenon quite challenging both in the phase. The algorithmic developments described here provide us a tool to study these questions. In
a later section, we illustrate this by providing strong computational evidence for a power-law intermediate temperature phase associated with the two-step melting of columnar order.

\subsection{TFIM with multi-spin interactions on the square lattice.}
\begin{figure}[t]
  \includegraphics[width=\columnwidth]{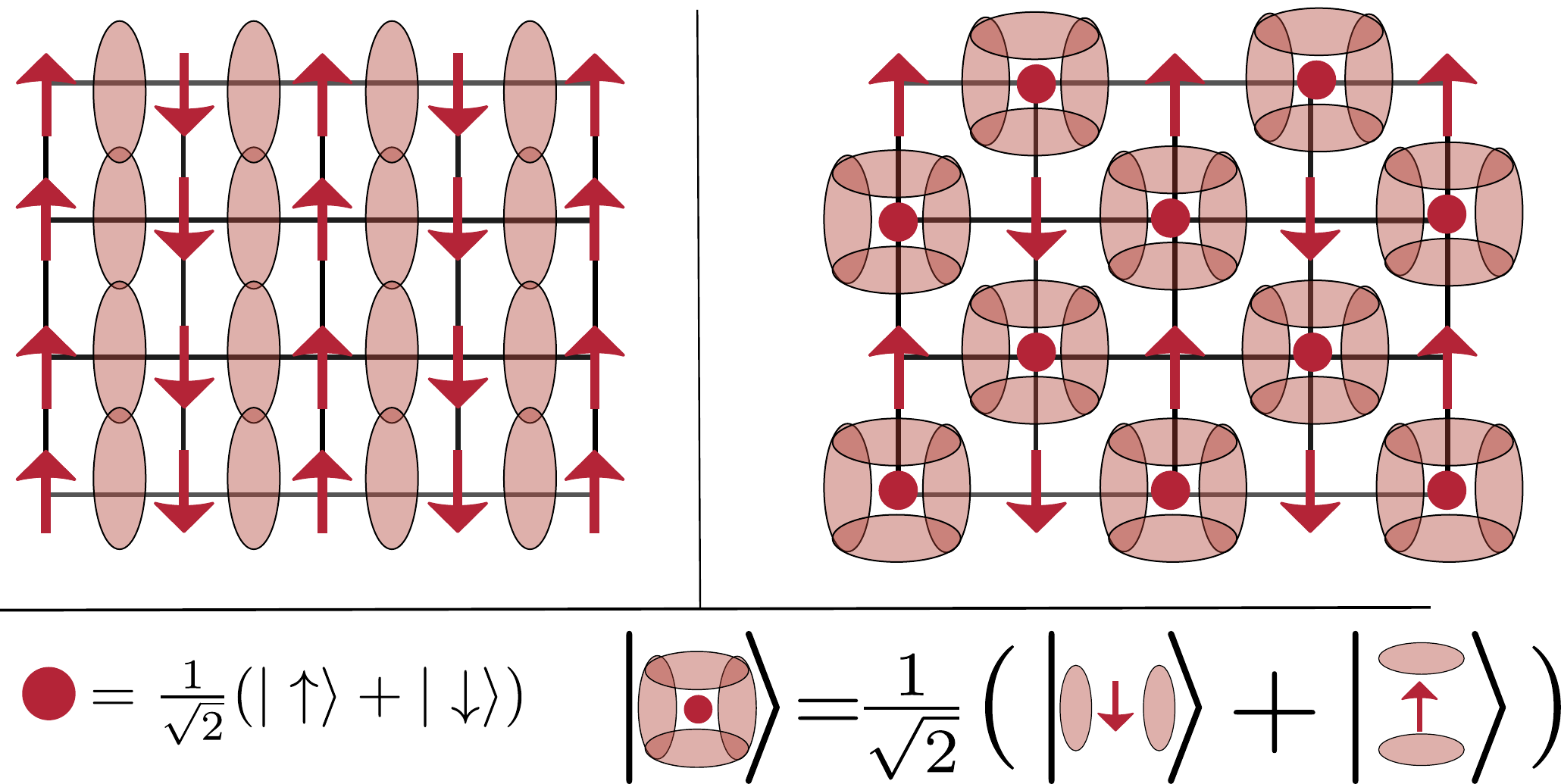}
  \caption{\label{EFIM_cartoon} Two natural ordered states in the quantum loop limit of the transverse field  Ising model with multispin interactions. The state on the left is the resonating plaquette state while the one on the right is the nematic state of loops. Each of these states
    is two fold degenerate. In the Ising model, each of these represents a four-fold degenerate state. The loop nematic corresponds to a striped Neel state while the resonating
    plaquette state corresponds to a state in which two diagonally opposite spins on
    every spatial plaquette are polarized in the $x$ direction by the transverse field and the
  others order antiferromagnetically.} 
\end{figure}
Two dimensional loop models have been of interest in statistical mechanics as lattice realizations of two dimensional conformal fixed points.\cite{Blotte_Nienhuis,Henley,Kondev,Jacobsen} (Decorated) quantum loop models have also been studied in the context of topologically ordered states and symmetry protected topological states.\cite{Tu,Senthil,Savary,Bauer} Here, we consider a TFIM designed to be precisely dual to the even Ising gauge theory, which reduces in a certain limit to the square lattice quantum loop model. The Ising
spins in our transverse field Ising model have multispin interactions defined on the square plaquettes of the 
square lattice. With respect to the four-sublattice decomposition of the square
lattice shown in Fig.~\ref{fftfim_lat}, the Hamiltonian is given by
\begin{align}
    H_{\mathrm{TFIM}}=&J\sum_p(\sigma^{z}_{A}\sigma^{z}_{B}\sigma^{z}_{C}\sigma^{z}_{D} \\
    \nonumber &+\sigma^{z}_{A}\sigma^{z}_{C}  +\sigma^{z}_{B}\sigma^{z}_{D} )_{p} -h\sum_{i}\sigma^x_i .
  \label{egt_decomp}
\end{align}
The index $p$ labels the elementary square plaquettes of the lattice, and the subscripts $A$, $B$, $C$ 
and $D$ denote the four sites belonging to the plaquette $p$ according to the 4-sublattice decomposition
of the square lattice shown in Fig.~\ref{fftfim_lat}. The index $i$ labels the sites of the lattice.

{\em Defining} a frustrated bond to correspond to a pair of antiparallel nearest neighbour spins, we see that the Ising energy of our model is minimized by configurations in which
each square plaquette has precisely two frustrated bonds. Higher Ising energy configurations
also have some plaquettes with zero or four frustrated bonds. Thus, the minimum
Ising energy configurations are macroscopically degenerate.

On the dual square lattice, we may again place a dimer on every dual link that crosses
a frustrated bond. Minimum Ising energy configurations now correspond to dimer
configurations in which each dual lattice site touched by precisely two dimers. Higher energy configurations also have some sites touched by zero or four dimers. Defining dual Ising
variables $\tau^x=-1$ on dual links covered by a dimer (and $\tau^x=+1$ on
empty dual links), we see that this is the configuration space of the even Ising gauge theory, with the Gauss law constraint requiring the product of $\tau^x$ over all dual links meeting at a dual site to be $+1$. As in the odd Ising
gauge theory case, the transverse field
term gives rise, in the dual language of the even gauge theory, to an energy
cost of $2\Gamma$ for a vison on a dual spatial plaquette.

However, this even Ising gauge theory is not the familiar gauge theory dual
to the ferromagnetic transverse field Ising model. In that case, the Ising
exchange term corresponds in gauge theory language to a uniform transverse
field that couples to $\tau^x$ and tries to align all the $\tau$ spins to point
along the $+x$ axis in $\tau$ space. In the present case, the Ising energy
corresponds to a sum of bilinears in $\tau^x$. Thus, this gauge theory
has the same vison energetics as the usual even Ising gauge theory, but
different dynamics.

\begin{figure}[t]
  \includegraphics[width=\columnwidth]{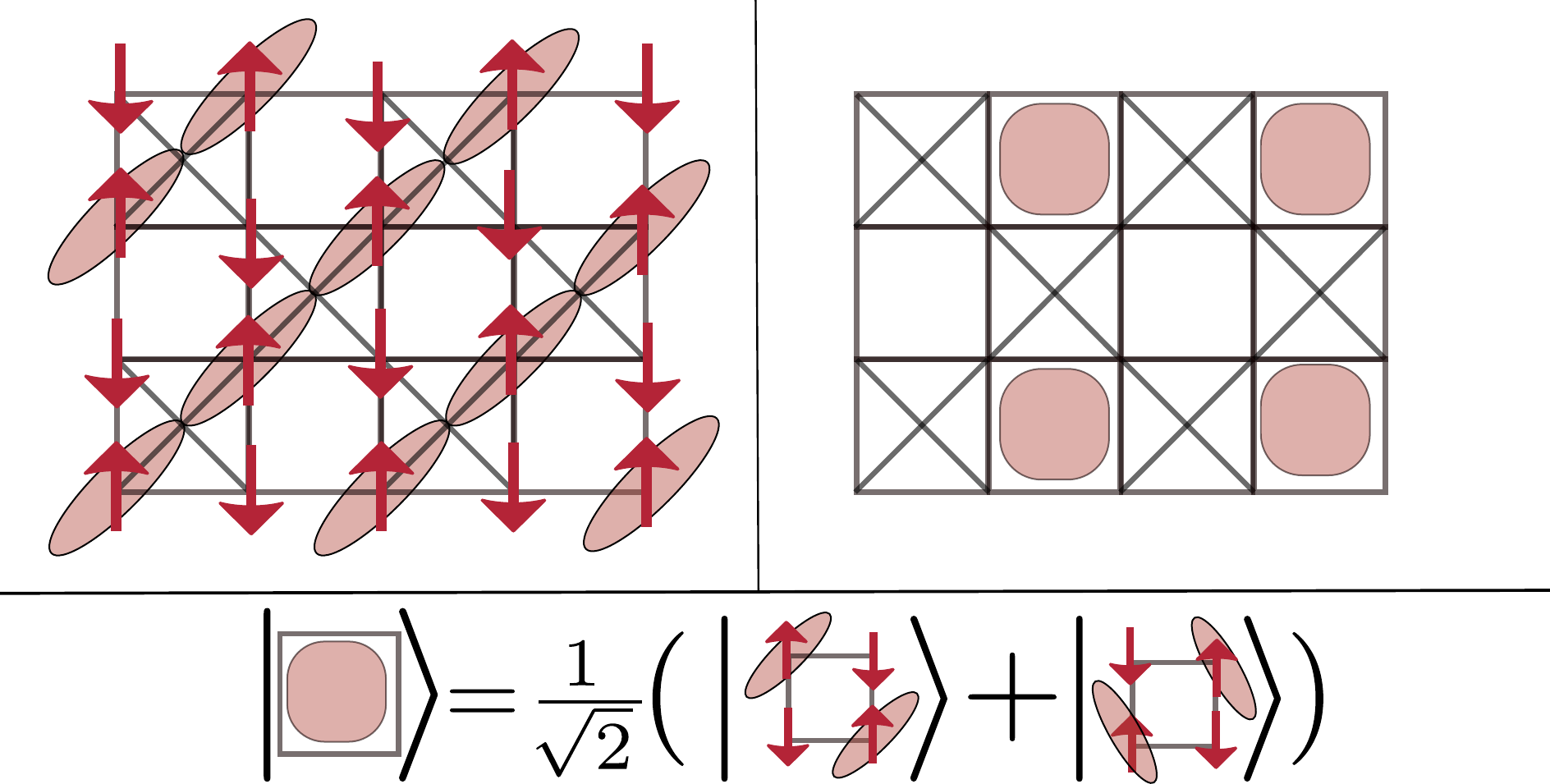}
  \caption{\label{quantumloopcartoons} Two natural ordered states in the low temperature limit of the transverse field Ising model on the planar pyrochlore lattice. The figure on
    the left is the Neel state in Ising language, which corresponds to the nematic state in the quantum loop language, while the
    one on the right corresponds to the resonating plaquette state. Both states are
  two fold degenerate in the Ising model.} 
\end{figure}

In the limit in which the Ising energy $J$ dominates over
the transverse field in the Ising model, it reduces to leading order in degenerate perturbation to the square lattice quantum loop model, with the usual
ring-exchange dynamics arising as the first perturbative effect. At larger values of
transverse fields, other terms appear in the effective Hamiltonian, including
some potential energy terms. For small
transverse fields, two natural ordering patterns are immediately suggested
by the form of the leading order effective Hamiltonian in this limit.
These are displayed in Fig.~\ref{quantumloopcartoons}. Since dimers correspond to frustrated bonds, the two fold symmetry breaking that characterizes these loop states implies four fold symmetry breaking for the Ising spins $\sigma^z$. 

In the limit of small quantum fluctuations (compared to the exchange or potential energy), several other systems reduce to the quantum loop model. These include the spin-1/2 XXZ antiferromagnet on the planar pyrochlore lattice,\cite{Shannon_Misguich_Penc}
and  the transverse field Ising antiferromagnet on the planar pyrochlore lattice (which
we discuss separately in a later section).\cite{Henry_Roscilde}  Exact diagonalization
of small systems suggest that the ground state in
this quantum loop model limit of the XXZ model is the resonating plaquette state.\cite{Shannon_Misguich_Penc} This is consistent with a direct diffusion Monte Carlo study of a quantum loop model defined in the language of arrow configurations of the classical six-vertex model.\cite{Syljuasen_Chakravarty} On the other hand, a direct numerical study of
this regime in the planar pyrochlore case could not find direct evidence for this state in the temperature range accessible to the numerics, whereas larger values of the
transverse field were found to favour a Neel state for the spins, that corresponds to a columnar ordered state of the loop model.\cite{Henry_Roscilde} 
Finally, we note that a recent numerical
work\cite{Banerjee_etal_JStat} on a three-dimensional classical statistical mechanics model (whose $\tau$ continuum limit yields a generalized quantum loop model) also finds a plaquette ordered phase and another nematic phase in which loops form parallel lines (see Fig.~\ref{quantumloopcartoons} for cartoons of both these phases).

Both these states correspond to a two-fold symmetry breaking in the quantum loop model.
In our transverse field Ising model, this corresponds to four-fold symmetry breaking since
the dimers represent bond-energies of the spin model.
As we demonstrate below, the methods developed here hold promise for a more direct quantum Monte Carlo study of the low temperature behaviour of the quantum loop model,
including a transition from a resonating plaquette state to the nematic state.
Additionally, our spin model allows for a direct study of this nontrivial even gauge theory
beyond the quantum loop model limit.

\subsection{TFIM on the pyrochlore lattice.}
\begin{figure}[t]
  \includegraphics[width=\columnwidth]{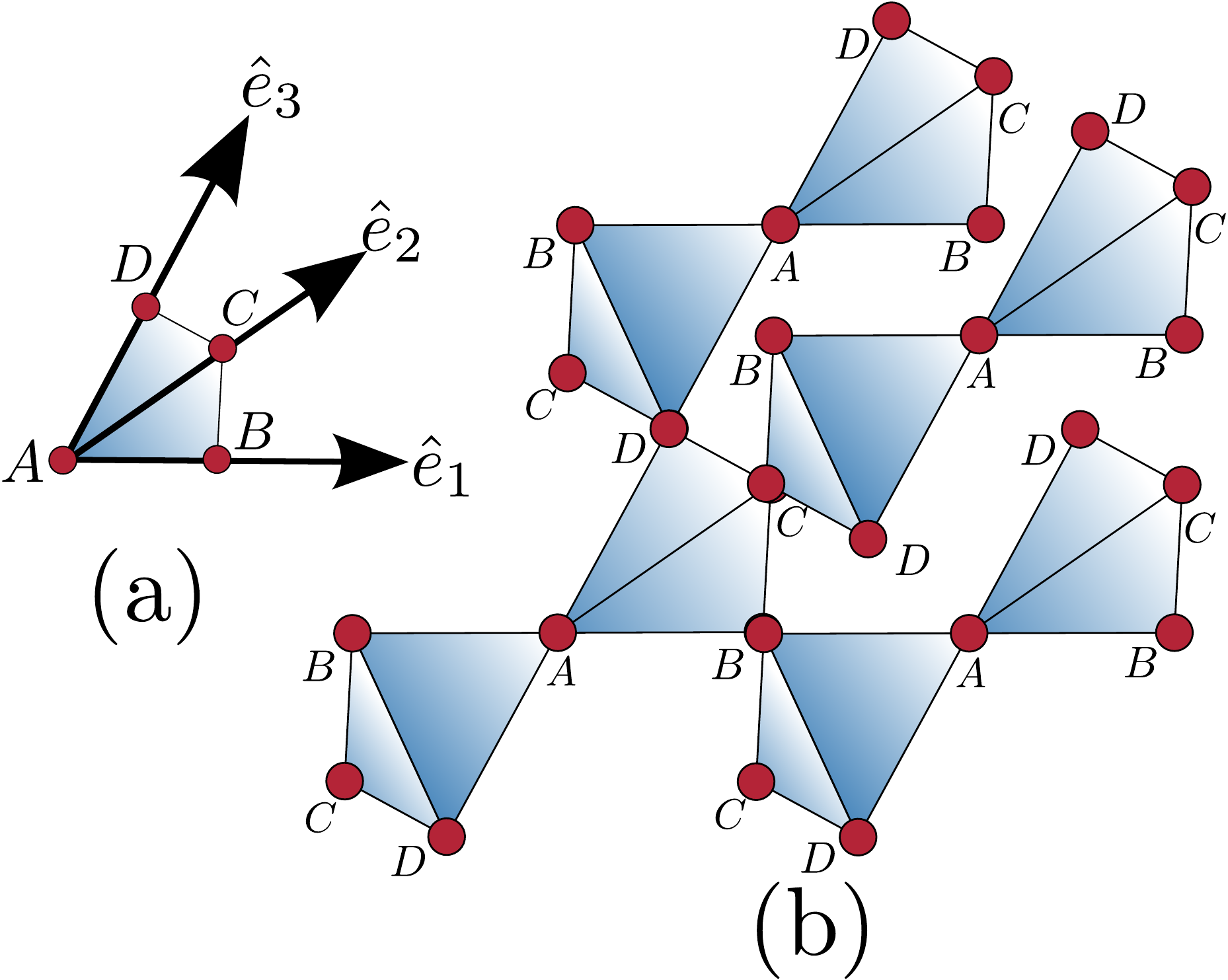}
  \caption{\label{pyrlattice}(a) The four site basis for the pyrochlore lattice (which defines the four sublattices of the full lattice) and the three lattice translation vectors $\hat{e}_1$,$\hat{e}_2$ and $\hat{e}_3$ that generate the lattice. (b) The pyrochlore lattice of corner sharing tetrahedra, generated by translating this basis of sites along linear combinations of the three lattice translation vectors. The edges of blue tetrahedra denote antiferromagnetic interactions
  between spins on the vertices.}
\end{figure}
\begin{figure}[t]
  \includegraphics[width=\columnwidth]{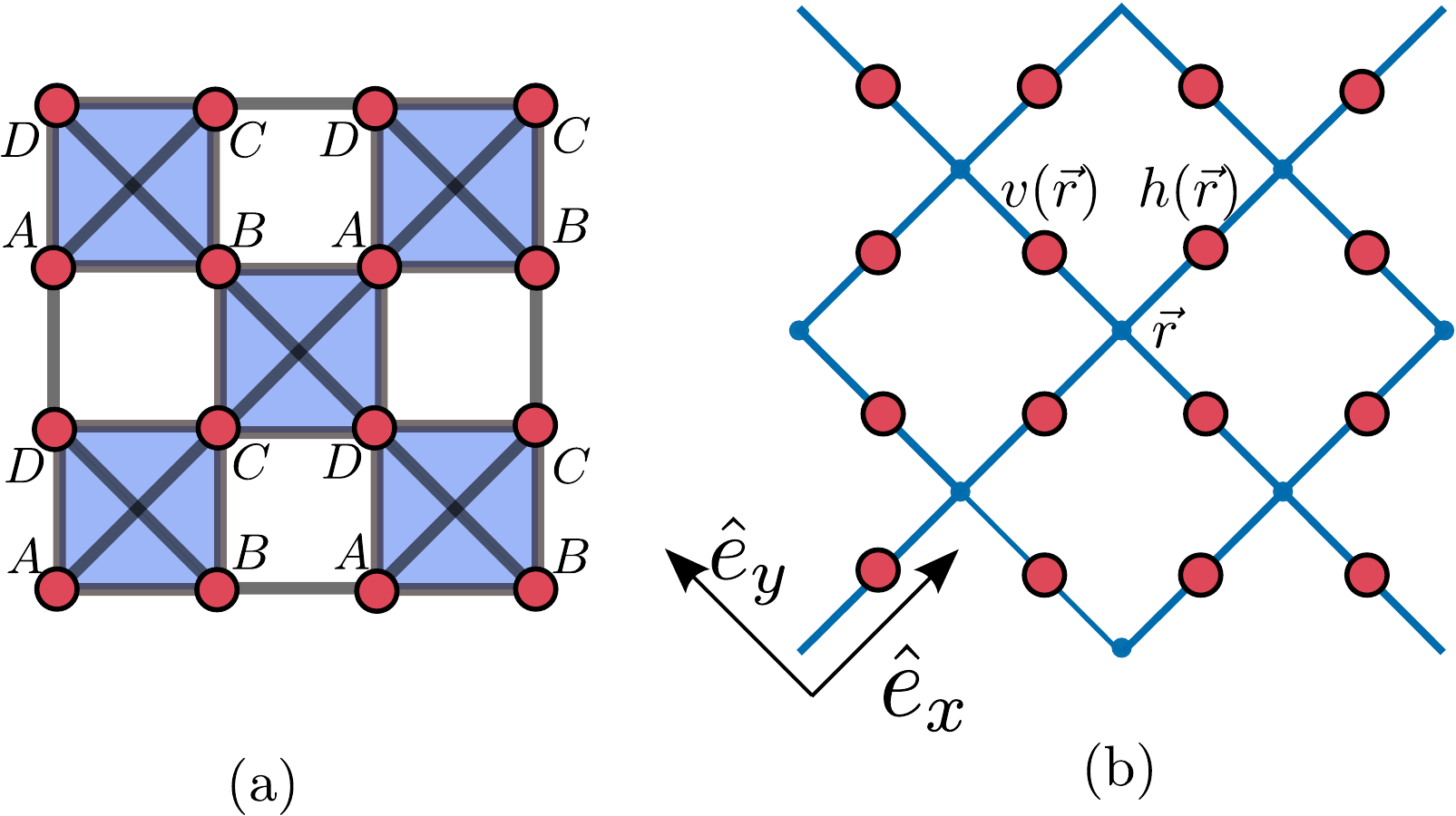}
  \caption{\label{planpyr_lat}(a) The planar pyrochlore lattice (also known as the square lattice with crossings). The black lines denote antiferromagnetic couplings between spins on the lattice sites, and the labels indicate the four sublattice structure. (b) The planar pyrochlore sites correspond centers of links of a square lattice. The sites of this square lattice sites are centers of the blue plaquettes (planar tetrahedra) of (a).  The spins on the four links touching a site $\vec{r}$ of this square lattice are the part of the planar tetrahedron labeled by $\vec{r}$. The links are labelled $v(\vec{r})$ and $h(\vec{r})$ as shown.} 
\end{figure}
The underlying Bravais lattice of the pyrochlore lattice is generated by the unit vectors $\hat{e}_1$, $\hat{e}_2$ and $\hat{e}_3$, which point along the edges of a regular
tetrahedron from one of its vertices. The pyrochlore lattice has a $4$ site basis, with the coordinates of the four sites being given by the displacement vectors $0$, $\hat{e}_1/2$, $\hat{e}_2/2$ and $\hat{e}_3/2$. This is illustrated in Fig.~\ref{pyrlattice}.

In the transverse field Ising antiferromagnet on the pyrochlore lattice, all nearest neighbor links between all sites host antiferromagnetic Ising-exchange interactions. The Hamiltonian is given by
\begin{align}
  \nonumber H_{\mathrm{TFIM}}=&J\sum_T(\sigma^{z}_{A}+\sigma^{z}_{B}+\sigma^{z}_{C} +\sigma^{z}_{D})^2_{T}\\
    &-h\sum_{i}\sigma^x_i .
\end{align}
Here, $T$ labels the tetrahedra making up the pyrochlore lattice, while the subscripts $A$, $B$, $C$ 
and $D$ denote the four sites belonging to the tetrahedron $T$ according to the 4-sublattice decomposition shown in Fig.~\ref{pyrlattice}. The index $i$ labels all sites of the lattice.

The Ising exchange energy of this model provides a good starting point
for the study of the physics of the classical spin ice compounds such
as Holmium Titanate.\cite{spinicereview} Other materials, in which
quantum effects are significant, are described by more complicated variants
of the simple transverse field Ising antiferromagnet that we focus on.

Interest in this class of systems arises in part from the observation\cite{Hermele_Balents_Fisher} that the ground state
in the presence of small (compared to the Ising exchange scale) but nonzero quantum fluctuations is likely to be a realization of a deconfined Coulomb
liquid phase. Direct numerical evidence of a Coulomb liquid state\cite{Banerjee_etal_PRL} came initially from quantum Monte Carlo studies of the $S=1/2$ XXZ model with strong
antiferromagnetic Ising exchange and weak ferromagnetic transverse exchange
couplings on the pyrochlore lattice. In spite of being the simplest
system expected to host a Coulomb liquid state, the transverse field Ising
antiferromagnet on the pyrochlore lattice has remained largely out of the reach of detailed and large-scale quantum Monte Carlo studies due to computational
difficulties. While recent work has mapped out the extent of the Coulomb liquid
phase using series-expansion methods\cite{Roechner_Balents}, and located the putative first-order transition using Quantum Monte Carlo methods,\cite{Emont_Wessel} a detailed quantum Monte Carlo characterization of the Coulomb liquid phase itself has not been attempted. We will see below that this is challenging due to computational difficulties associated with the accurate measurement of the
structure factor of $\sigma^z$ in conventional QMC algorithms, and that our method holds promise in this regard.

\subsection{TFIM on the planar pyrochlore lattice.}
The planar pyrochlore or the 2D pyrochlore is similarly defined as a square Bravais lattice, with a 2-site basis, described by the coordinates $\hat{e}_x/2$ and $\hat{e}_y/2$ as shown in Fig.~\ref{planpyr_lat}. Every shaded plaquette in Fig.~\ref{planpyr_lat} hosts antiferromagnetic interactions
between all sites on the plaquette, and can be thought of as a planar version of a tetrahedron.
The Hamiltonian is given by
\begin{align}
  \nonumber H_{\mathrm{TFIM}}=&J\sum_p(\sigma^{z}_{A}+\sigma^{z}_{B}+\sigma^{z}_{C} +\sigma^{z}_{D})^2_{p}\\
    &-h\sum_{i}\sigma^x_i ,
\end{align}
where $p$ denotes the \emph{shaded plaquettes} illustrated in Fig.~\ref{planpyr_lat}. The subscripts $A$, $B$, $C$ and $D$ denote the sites
belonging to the plaquette $p$ according to the 4-sublattice decomposition of the square lattice shown in Fig.~\ref{planpyr_lat}.

The Ising exchange energy is clearly minimized by configurations in which each
planar tetrahedron has two sites with $\sigma^z = +1$ and two sites
with $\sigma^z = -1$. Representing the up spins by dimers on the associated
square lattice (whose links host the Ising variables), we see that minimum
exchange energy configurations correspond to configurations of the loop
model in which each square lattice site is touched by exactly one loop (two dimers). Higher energy configurations contain some defects, {\em i.e.}
sites touched by three or even four dimers. For low values of $\Gamma$, 
the leading order effective Hamiltonian corresponds to a quantum loop
model with a weak ring exchange term that scales as $\Gamma^4/J^3$. Thus,
this effective Hamiltonian again favours the low temperature states
discussed earlier in the context of the quantum loop model limit of
the frustrated square lattice Ising model with multispin interactions.
There are two key differences:  First, the temperature scale associated with the ordering
is expected to be of order $\Gamma^4/J^3$ rather than $\Gamma$. Second, since dimers are now
up spins, the symmetry breaking in the spin model is two-fold in nature, described by an Ising-like
order parameter. As mentioned earlier, a direct numerical study of
this regime in the planar pyrochlore case could not find direct evidence for resonating plaquette state in the temperature range accessible to the numerics, whereas larger values of the
transverse field were found to favour a Neel state for the spins, that corresponds to a columnar ordered state of the loop model.\cite{Henry_Roscilde} The method presented
here could in principle be used to study the transition between these two states of
the transverse field Ising antiferromagnet on the planar pyrochlore lattice.

\section{The Stochastic Series Expansion.}
\label{review}

\begin{figure}[t]
  \includegraphics[width=5cm]{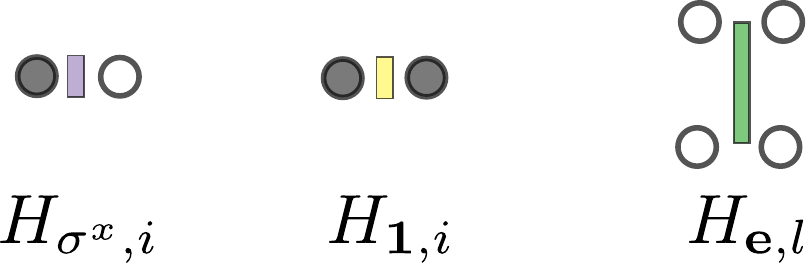}
  \caption{A graphical representation of operators in the SSE operator string takes the form of vertices with two legs corresponding to each site on which the operator acts.}
  \label{ssevertices}
\end{figure}

\begin{figure}[t]
  \includegraphics[width=\columnwidth]{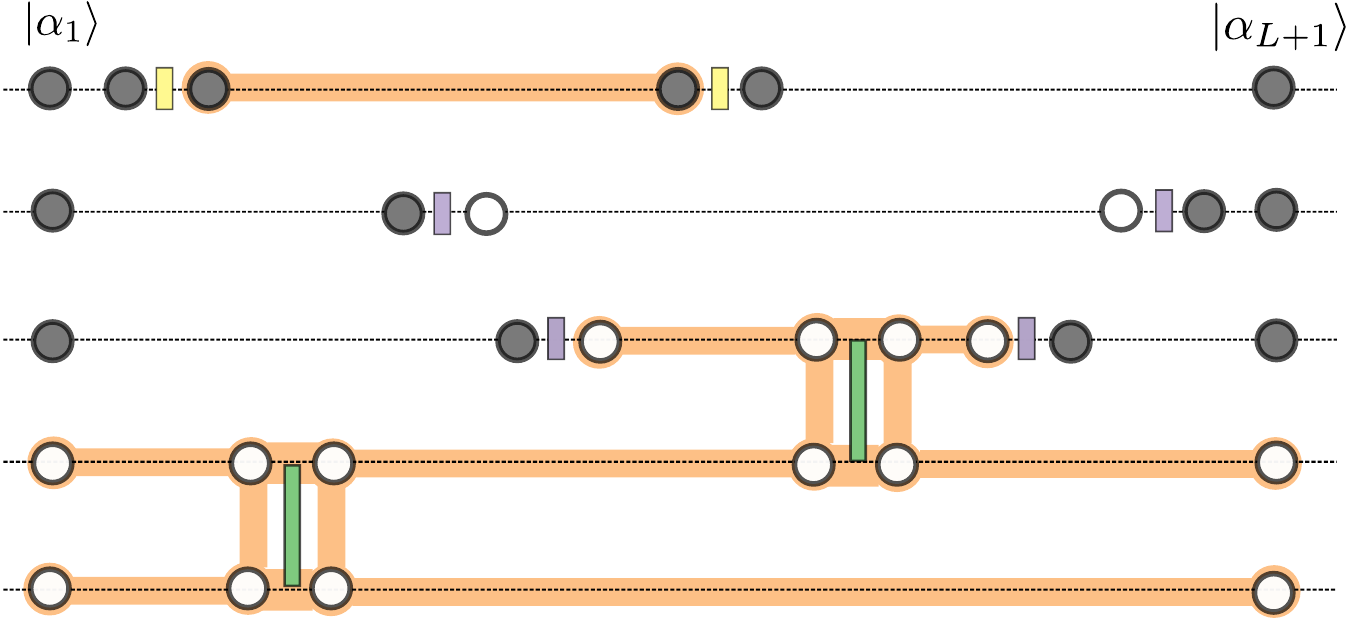}
  \caption{A graphical representation of the operators in a SSE operator string, with an illustration of a decomposition in terms of clusters according to the rules formulated in Ref~\onlinecite{Sandvik_TFIM}.}
  \label{opstring}
\end{figure}

The Stochastic Series Expansion (SSE) developed by Sandvik\cite{Sandvik_SSE} is a Quantum Monte Carlo method which involves sampling  terms from the high-temperature expansion of the partition function. Here, we summarize this
approach and describe Sandvik's quantum cluster algorithm\cite{Sandvik_TFIM} for the transverse field
Ising model. In the standard SSE approach, the Hamiltonian is first decomposed as a sum of transverse field terms living on the sites of a lattice  and Ising exchange terms living on links of the lattice :
\begin{equation}
  \begin{split}
    &H_{\mathrm{TFIM}}=-\sum_{l}H_{\mathbf{e},l} -\sum_{i}H_{\mathbf{1},i} -\sum_{i}H_{\sigma^{x},i},\\
    & H_{\mathbf{e},l}=|J_{1}|-J_{1}\sigma^{z}_{1(l)}\sigma^{z}_{2(l))},\\
    & H_{\mathbf{1},i}=\Gamma {\mathbf 1}_i,\\
    & H_{\sigma^{x},i}=\Gamma \sigma^{x}_{i}.
  \end{split}
  \label{Hdecomp}
\end{equation}
The subscript $l$ in $H_{\mathbf{e}.l}$ labels the link on which the corresponding exchange coupling acts. For these operators, $1(l)$ and $2(l)$ denote the sites connected by the link $l$. For $H_{\mathbf{1},i}$ and $H_{\sigma^x,i}$, $i$ labels the sites of the lattice. 
Constants have also been added to the Ising-exchange operators so that all eigenvalues of each of these operators is positive or zero.
Further, all three kinds of operators have a no-branching property in the basis $\{ |\alpha \rangle\}$ of $\sigma_z$ eigenstates, {\em i.e.} each kind of operator acting on a basis state gives
a state that is proportional to a single basis state, not a linear superposition of basis states.
The Taylor expansion of the partition function $Z=Tr(\exp(-\beta H))$ in the inverse temperature $\beta$ now reads
\begin{equation}
  Z=\sum^{\infty}_{n=0}\frac{\beta^{n}}{n!}\sum_{S_{n},|\alpha_1\rangle}\prod^{n}_{k=1}\langle \alpha_{k+1} |H_{a(k),b(k)}|\alpha_{k} \rangle \;.
  \label{SSEeqn}
\end{equation}
Here $S_{n}$ schematically labels all possible operator strings of length $n$ that act on an initial basis state $|\alpha_1\rangle$ to produce a final state proportional to $|\alpha_1\rangle$, $b(k)$ is either a site or a link and $a(k)$ is either $\sigma^{x}$ or $\mathbf{1}$ if
$b(k)$ is a site, and $a(k)= \mathbf{e}$ if $b(k)$ is a link. Note that the \emph{no-branching} property alluded to above ensures that $H_{a(k),b(k)}|\alpha_{k}\rangle \propto |\alpha_{k+1}\rangle$ for a unique choice of $|\alpha_{k+1}\rangle$, allowing the trace to be written in this form. The periodicty of the trace appearing in the partition function imposes the condition $|\alpha(1)\rangle =|\alpha(n+1)\rangle$. The operator string may be represented graphically
by depicting each operator as shown in Fig.~\ref{ssevertices} and linking them up
as shown in Fig.~\ref{opstring}.
Further, for practical computations it is convenient to use a fixed length representation, which fixes an upper cut-off $\mathcal L$ for the length of the operator string. This involves introducing $\mathcal{L}-n$ dummy identity operators (denoted by $H_{0,0}$) in all possible ways into an operator string of length $n$ to convert it into a padded operator string of length $\mathcal L$. There are $\mathcal{L} \choose \mathcal{L}-n$ such ways to do it, which converts equation ~\eqref{SSEeqn} into 
\begin{equation}
  Z=\frac{1}{{\mathcal L}!}\sum_{S_{{\mathcal L}}}\beta^{n}({\mathcal L}-n)!\prod^{{\mathcal L}}_{k=1}\langle \alpha_{k+1} |H_{a(k),b(k)}|\alpha_k\rangle \; ,
  \label{LSSE}
\end{equation}
where $a(k)=0$, $(b(k)=0$ is now an additional possibility allowed to incorporate the presence of dummy operators $H_{0,0}$.
The periodicity of the trace now imposes $|\alpha(1)\rangle =|\alpha(\mathcal{L}+1)\rangle$.
The Quantum Monte Carlo method now samples the configuration-space spanned by the operator strings $S_{\mathcal L}$ and the initial basis states $|\alpha_1\rangle$. This is accomplished by two updates, the \emph{diagonal update} which switches the identity ($H_{0,0}$) operators with diagonal operators while preserving detailed balance, and the \emph{quantum cluster update}, which switches the off-diagonal transverse-field operators $H_{\sigma^{x},i}=\Gamma \sigma^{x}_i$ with the diagonal operators $H_{\mathbf{1},i}=\Gamma \mathbf{1}_i$.


The \emph{diagonal update} involves passing through the operator string and replacing the identity operators $H_{0,0}$ with the diagonal operators ($H_{\mathbf{e},l}$ and $H_{\mathbf{1},i}$) and vice-versa. The probabilities are dictated by the requirement of satisfying detailed balance with respect to weights of the corresponding configurations in the operator string in Eq.~\eqref{LSSE}:
\begin{equation}
  \begin{split}
    & P(H_{0,0}\rightarrow  {\mathrm{diag.}} \; {\mathrm{operator}})=\mathrm{Min}\Big(1,\frac{\beta(N\Gamma+2|J|N_{links})}{{\mathcal L}-n}\Big),\\
    &  P( {\mathrm{diag.}} \; {\mathrm{operator}} \rightarrow H_{0,0})=\mathrm{Min}\Big(1,\frac{{\mathcal L}-n+1}{\beta(N\Gamma+2N_{\mathrm{links}}|J|)}\Big),
  \end{split}
  \label{diagup}
\end{equation}
where $N$ is the number of sites on the lattice and $N_{\rm{links}}$ is the number of links. On choosing to replace an identity operator $H_{0,0}$ by a diagonal operator, each $H_{\mathbf{1},i}$ operator is chosen with a probability $\Gamma/(N\Gamma+2|J_{1}|N_{links})$ while each $H_{\mathbf{e},l}$ is chosen with a probability $2|J_{1}|/(N\Gamma+2|J_{1}|N_{links})$. 
A particular Ising exchange operator might not be allowed at a particular value of $k$ if
the spin configuration $|\alpha_k\rangle$ corresponds to zero weight for that
particular exchange operator.  In that case, the replacement is aborted and the sweep is continued by attempting to update the next diagonal or dummy operator.

The \emph{cluster update} attempts to effect large scale non-local changes in the operator string by exchanging multiple off-diagonal transverse-field operators $H_{\sigma_{x},i}$ with the corresponding diagonal operators $H_{\mathbf{1},i}$ and vice-versa. In the process the spin configuration at a large number of ``time slices'' $k$ can also change. Since the $H_{\mathbf{1},i}$($\Gamma \mathbf{1},i_i$) and $H_{\sigma_{x},i}$($\Gamma \sigma^{x}_i$) operators appear with equal weights in the operator string, this leads to a \emph{microcanonical cluster update}. The rules for the cluster construction are best stated graphically, as in Fig.~\ref{opstring}: All legs of the operator vertex corresponding to an exchange operator are assigned
to the same cluster, while the two legs of operator vertices corresponding to the transverse
field operators (either $H_{\mathbf{1},i}$ or $H_{\sigma^x,i}$ are assigned a priori to two
  different clusters (which could however link up via a common exchange operator when
  the cluster constructions are completed). Thus, the transverse field operators are
  responsible for attenuating the growth of clusters, while the exchange operators lead
  to their spread and growth. In this sense, the cluster construction involves a link
  percolation process.

Flipping a cluster corresponds to flipping all spin states on all legs of the cluster. These changes in the basis
states $\alpha_k$ can lead to $H_{\mathbf{1},i}$ and $H_{\sigma_{x},i}$ operators being
exchanged with each other. These cluster construction rules can be used in two different ways:
a) The \emph{Swendsen-Wang} update\cite{Swendsen_Wang} scheme, in which one divides  up all legs of the operator string into mutually non-overlapping clusters and then flips each cluster with a probability $P_{flip}=1/2$, or 
b) The \emph{Wolff} update\cite{Wolff} scheme, in which one 
starts growing a cluster from a randomly selected leg, and flips that one cluster with probability $P_{flip}=1$. While the original algorithm of Ref.~\onlinecite{Sandvik_TFIM} employed
the Swendsen-Wang approach, we will have occasion to use both
approaches in the methods presented here.

\section{Need for premarked motifs and one-dimensional clusters}
\label{needmicrocan}
Frustration results in a large number of  {\em inequivalent} (unrelated by any global symmetry) of the classical
exchange energy. Even in the absence of a transverse field term, thermal fluctuations
can lift this degeneracy, since different minima have a weight controlled
at nonzero temperature by the entropy of nearby states that can be accessed by thermal fluctuations. Quantum
fluctuations resulting from a transverse field are also responsible for a similar
selection mechanism, since the zero point energy of quantum fluctuations about
different minima is typically quite different. In such frustrated models, the usual cluster algorithms that
work well in the ferromagnetic case typically fail because the corresponding clusters
percolate and freeze, rendering them nearly useless. This reflects the fact that
the usual algorithms do not incorporate the physics of these thermal and quantum
fluctuations about a large number of symmetry-inequivalent classical minima of the exchange energy.

This has been noted previously, and attempts have been made to alleviate the problems in the context of 
both classical\cite{KBD_PRL,Kandel_Domany,KBD_PRB,Zhang_Yang,Coddington_Han} and quantum systems.\cite{Louis_Gros,Kedar_Dariush, Henry_Roscilde, Biswas_Rakala_Damle}
As has been noted in some of these earlier works, the root of the difficulty is that a decomposition of the exchange energy into
terms living on links is inappropriate in the frustrated case, since it does not contain enough information about the geometry of the lattice which causes competition between different
exchange coupling and leads to a multiplicity of minima for the exchange energy. To incorporate the physics of
this competition, it is therefore necessary, at a minimum, to decompose the Hamiltonian into the {\em natural elementary plaquettes} that capture the minimal geometric information that determines the nature of the minimum energy configurations of the corresponding classical system: triangles for the triangular lattice, frustrated squares for the fully frustrated Ising model on the square
lattice, and tetrahedra for the pyrochlore lattices. In other words, it does not in general suffice to simply  follow Ref.~\onlinecite{Biswas_Rakala_Damle} and decompose the Hamiltonian into triangular plaquettes in problems where the natural geometric unit is larger, for instance a tetrahedron in the case of the pyrochlore lattice---this was for instance also noticed in a recent test\cite{Emont_Wessel} of the method of Ref.~\onlinecite{Biswas_Rakala_Damle} on the pyrochlore lattice. Indeed, for the pyrochlore lattice, this is not even guaranteed to equilibriate well, making finer distinctions in terms of autocorrelation times rather moot.

Armed with these decompositons, previous authors have utilized the general Kandel-BenAv-Domany framework\cite{KBD_PRL, Kandel_Domany,KBD_PRB} to come up with fruitful cluster algorithms for classical Ising models.  Such a decomposition also
leads to efficient \emph{loop updates} for frustrated quantum spin models with conserved $S_z$.\cite{Louis_Gros,Kedar_Dariush} However, it does {\em not}  in general suffice to merely decompose the Hamiltonian using the natural plaquette decomposition for a given frustrated lattice to obtain more efficient quantum cluster updates. For instance, merely decomposing the pyrochlore lattice transverse field Ising model into terms living on tetrahedra does {\em not} yield an efficient quantum cluster update. This is at the heart of the performance difficulties noted in a recent study on the pyrochlore lattice.\cite{Emont_Wessel}

What then is the other ingredient responsible for the efficiency of the plaquette percolation based quantum cluster update developed earlier\cite{Biswas_Rakala_Damle} for the transverse field Ising antiferromagnet on the triangular lattice? As already noted in Ref.~\onlinecite{Biswas_Rakala_Damle}, it is equally important to maintain a certain consistency of the cluster decomposition rules as a function of ``imaginary time''
 (more precisely, along the length of the entire operator string). More precisely, \emph{if the same cluster-decomposition rule is not chosen whenever possible for all vertices that live on the same spatial plaquette}, the resulting clusters typically percolate and freeze, eliminating
all hoped-for advantages of starting with a plaquette-based decomposition of the Hamiltonian.

In our earlier work, this consistency in imaginary time was hardwired into our procedure
by first choosing a so-called ``privileged site'' on each triangular plaquette of the spatial lattice in some manner. Once this favoured site was marked, this premarking scheme determined the rule by which the legs of all diagonal operators living on any plaquette were decomposed into clusters.
The generalization to the models of interest here is now obvious: One
needs to pick a set of motifs, and some strategy for choosing and imprinting one of them on each spatial plaquette of the lattice. These premarked
motifs should fix, in a useful way, the rule for the cluster decomposition of all diagonal operators living on a given spatial plaquette. This will ensure consistency of the cluster decomposition rule along the operator string, and help produce a broad distribution of cluster sizes.
This is the first important takeaway from the success of Ref.~\onlinecite{Biswas_Rakala_Damle} on the triangular lattice.

Apart from using the notion of premarked privileged sites to ensure this consistency in imaginary time, the approach of Ref.~\onlinecite{Biswas_Rakala_Damle} had another feature that contributed to its efficiency: Included among the clusters this approach makes are single-spin flip
clusters, {\em i.e.} clusters that only affect the state of a single site. Such ``one-dimensional'' clusters are the QMC analog of the Metropolis single-spin flip moves in a classical Ising model, and are useful in improving the equilibriation and autocorrelation of off-diagonal observables such as $\sigma^x$. In Ref.~\onlinecite{Biswas_Rakala_Damle}, such one-dimensional clusters were automatically produced in the triangular lattice problem by the quantum cluster approach developed there. However, it is clear that this does not generalize: In other cases, where the minimum exchange energy configurations are characterized by a constraint on the total Ising spin of the elementary plaquette (for instance, the zero net spin condition on each tetrahedron in the pyrochlore Ising antiferromagnets), such a one-dimensional cluster simply cannot be grown within the {\em microcanonical} framework used in the quantum cluster update. In other words, the appropriate generalization of this feature of the algorithm of Ref.~\onlinecite{Biswas_Rakala_Damle} is a {\em canonical} update scheme that makes one dimensional clusters. This is the other takeaway from Ref.~\onlinecite{Biswas_Rakala_Damle}.

Below, we use these lessons to design efficient quantum cluster updates for the other interesting problems reviewed in the foregoing. Our strategy here is two-pronged. First, we generalize the notion of privileged sites to specify premarked motifs that are imprinted on the spatial lattice and guide the microcanonical cluster construction procedure. Second, we develop appropriate {\em canonical} cluster updates that construct one-dimensional single-spin clusters whenever the microcanonical cluster construction does not produce such single-spin clusters due to energetic considerations.

\begin{figure}[t]
  \includegraphics[width=8.5cm]{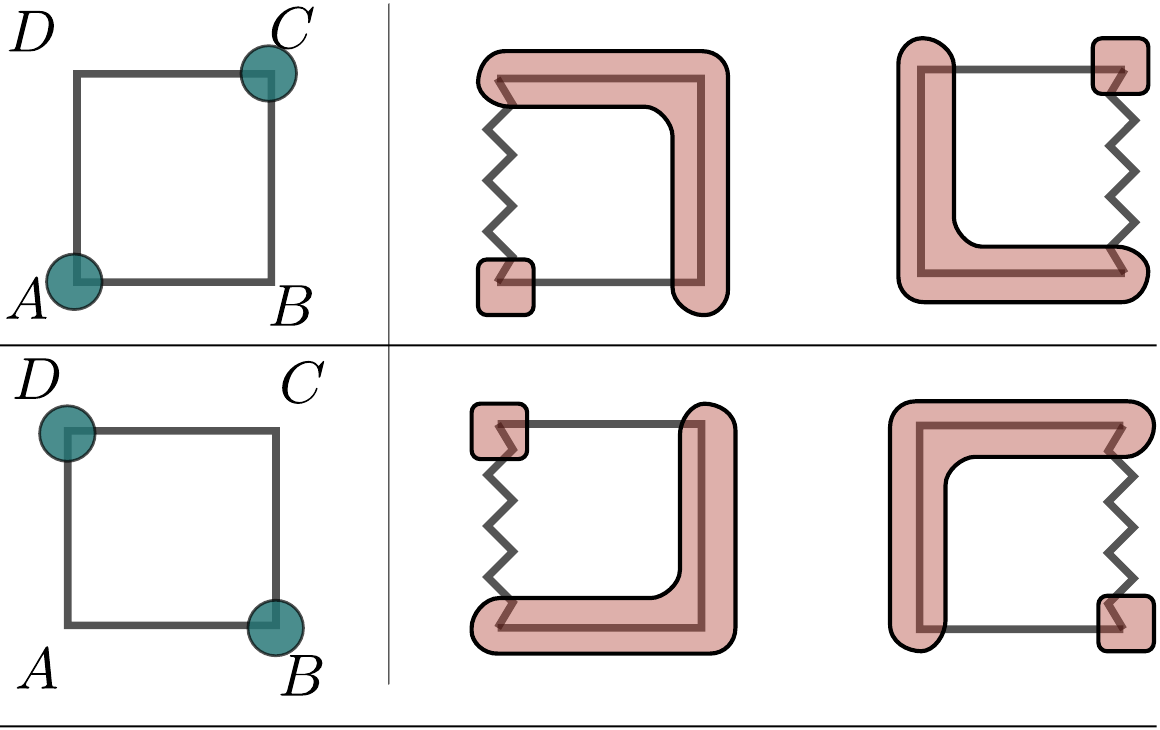}
  \caption{\label{fftfim_rules} The motifs for the fully frustrated transverse field Ising model (FFTFIM)  on the square lattice correspond to a pair of diagonally opposite sites on a spatial plaquette. There are thus two possible motifs on each spatial plaquette, one which
  picks out an A and a C sublattice site diagonally opposite to each other, and another which picks out a B and a D
  sublattice site diagonally opposite to each other. The cluster decomposition rules corresponding to each
  motif are shown on the right. In the figure, frustrated bonds of
  the spin configuration are denoted by zigzag lines, while unfrustrated bonds are denoted by straight lines.}
\end{figure}

\section{Microcanonical cluster update}
\label{microcanonical}
\subsection{Choice of motifs and associated cluster-decomposition rules}

Since the Ising-exchange energy in the Hamiltonian is now decomposed into elementary plaquettes instead of links,
the SSE operator string now consists of diagonal plaquette operators corresponding to the Ising-exchange, as well as the single
site operators of Sec.~\ref{review}. As in the standard link decomposition approach, a constant is added to the Ising exchange energy to ensure that each plaquette operator has matrix elements that are positive or zero. The cluster decomposition rules for the single site
operator vertices corresponding to $H_{\mathbf{1},i}$ 
and the $H_{\sigma^{x},i}$ remain the same as in the standard approach.
However, there is now some freedom in choosing the decomposition rule for
assigning different legs of a plaquette operator to clusters. This is taken care
of by our choice of motifs and premarking scheme in each case.

\subsubsection{Fully frustrated TFIM on the square lattice.} 
As alluded to in the preceding discussion, in the fully frustrated TFIM on the square lattice, the Ising-exchange operator is 
decomposed into elementary square plaquettes.
For the gauge choice described in Sec.~\ref{models}, the corresponding decomposition of
the Hamiltonian is
\begin{align}
    H_{\mathrm{TFIM}}=&-\sum_{p}H_{\Box,p} -\sum_{i}H_{\mathbf{1},i} -\sum_{i}H_{\sigma^{x},i}\\
  \nonumber   H_{\Box,p}=& J-\frac{J}{2}(\sigma^{z}_{D}\sigma^{z}_{A} -\sigma^{z}_{A}\sigma^{z}_{B} -\sigma^{z}_{B}\sigma^{z}_{C}  \\
  \nonumber &-\sigma^{z}_{C}\sigma^{z}_{D} )_{p}.
  \label{fftfim_decomp}
\end{align}
$p$ labels the square plaquettes of the lattice, and $A, B, C$ and $D$ denote
the four sites making up the plaquette, according to the 4-sublattice decomposition shown in Fig.~\ref{fftfim_lat}. The minimum energy eigenstates of $H_{\Box,p}$ corresponds to ones in which 
interactions in three out of the four links of the plaquette $p$ are satisfied, and have an energy of $2J$. There are $8$ such spin states. 
The other configurations, which correspond to 
only one link  of the plaquette hosting satisfied interactions, have weight zero, and the
corresponding diagonal plaquette operators do not appear in the operator string. Therefore, we do not consider them further.

Thus, all Ising-exchange vertices have a configuration in which one bond is frustrated.
A useful choice of motif consists of one of the two pairs of diagonally opposite sites of the square plaquette. There are thus two choices of motif for any given spatial plaquette.
For either choice, the frustrated bond must touch one of the sites that make up the motif. The corresponding decomposition rule is that we put the legs corresponding to this site into an a priori different cluster, while the remaining six legs make up the other cluster (of course
these two clusters can merge at a future step of the cluster construction).
These motifs and the corresponding cluster decomposition rules are illustrated in Fig.~\ref{fftfim_rules}.
Note that one might have naively thought that a different cluster decomposition rule, in which legs belonging to two of the sites form one cluster and the other four legs form the other
cluster, would be more optimal from the point of view of obtaining a broad distribution of
cluster sizes. However, if one chooses this kind of decomposition, there does not seem to be any simple way of ensuring consistency of the decomposition among all operators that live on a given spatial plaquette. Therefore, while this cluster-decomposition
works for the classical fully frustrated Ising model on a square lattice,\cite{Kandel_Domany,KBD_PRB, Coddington_Han} it fares badly (compared to the decomposition guided by
our premarked motifs) in the quantum case.

\subsubsection{TFIM on the pyrochlore and planar pyrochlore lattices.}
For the pyrochlore lattices, elementary plaquettes are the tetrahedra of four sites for the 3D pyrochlore lattice,
and the square plaquettes with crossings for the 2D pyrochlore lattice. The Hamiltonian is decomposed as
\begin{align}
    H_{\mathrm{TFIM}}=&-\sum_{p}H_{\tetr,p} -\sum_{i}H_{\mathbf{1},i} -\sum_{i}H_{\sigma^{x},i}\\
    \nonumber   H_{\tetr,p}=&8J-\frac{J}{2}(\sigma^{z}_{A}+\sigma^{z}_{B} +\sigma^{z}_{C} +\sigma^{z}_{D} )_p^{2},
  \label{pyr_decomp}
\end{align}
where $p$ labels the tetrahedra (shaded square plaquettes with crossings) for the 3D(2D) pyrochlore lattice, as illustrated in Fig.~\ref{pyrlattice}(Fig.~\ref{planpyr_lat}). The highest energy
configurations of a plaquette
correspond to having all four  spins pointing in the same direction, and these have an eigenvalue of $0$ for  $H_{\tetr,p}$. Therefore the corresponding
vertices do not appear in the operator string. 
Next, there are eigenstates of $H_{\tetr,p}$ with an eigenvalue of $6J$, which correspond to three spins being aligned to each other, and the fourth anti-aligned to these three in a tetrahedron.
Finally there are eigenstates with an eigenvalue of $8J$ and these correspond to the \emph{ice-rule} states, {\em i.e.} with two spins pointing up,
and the the other two pointing down.

For the vertices with weight $6J$, microcanonical decompositons involve separating out four legs corresponding to two antiparallel spins into a separate 
cluster, leaving behind another cluster of four legs corresponding to two parallel spins (as explained earlier, these two clusters can of course merge at a future step in the cluster construction). 
For the ice rule vertices with weight $8J$, a microcanonical decomposition corresponds to splitting
the eight legs into two clusters, each corresponding to a pair of antialigned spins (again with
the proviso that these two a priori different clusters could merge at a future step).

\begin{figure}[t]
  \includegraphics[width=8.5cm]{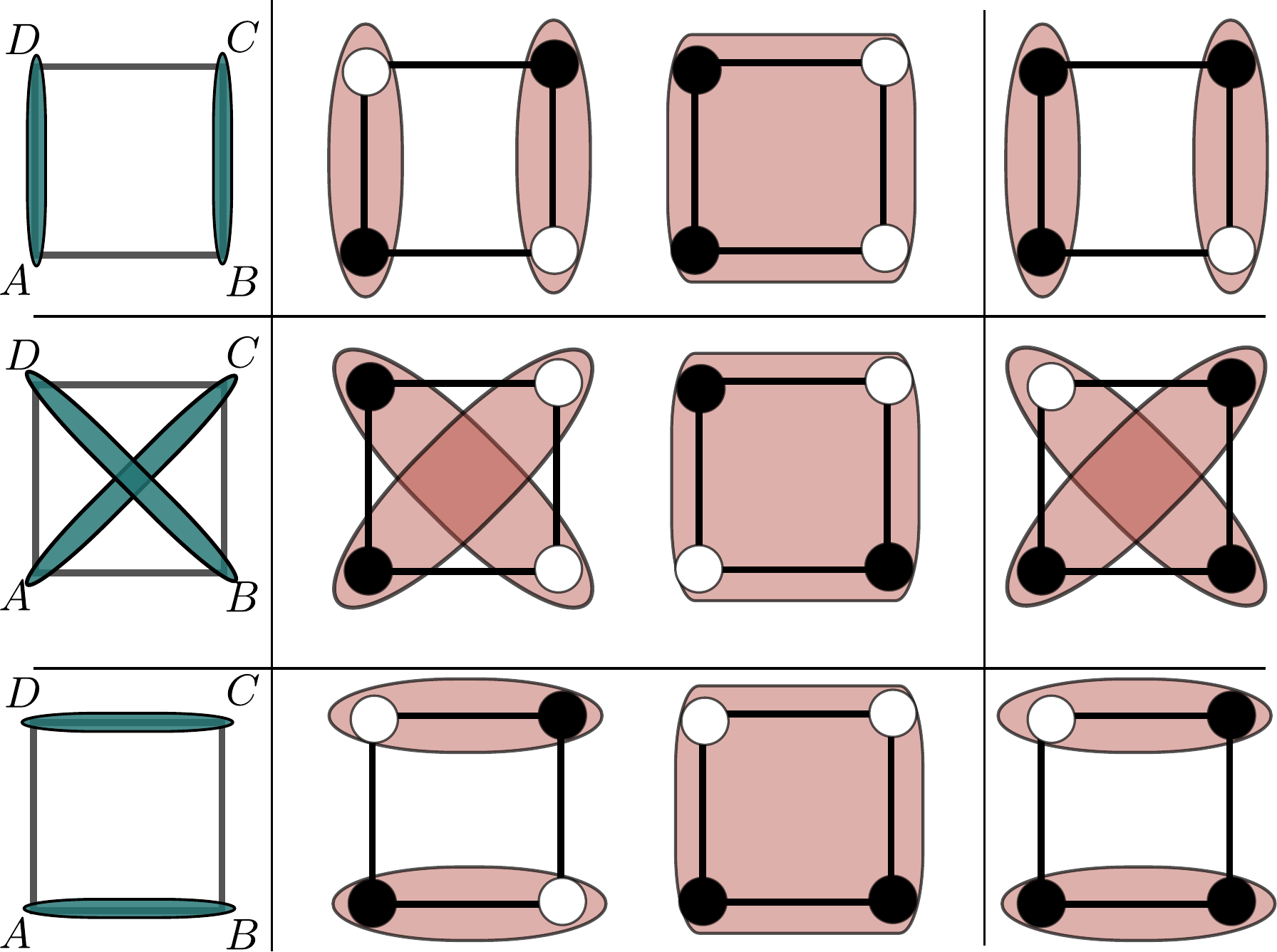}
  \caption{\label{pyr_rules} The three possible motifs for the pyrochlore and planar pyrochlore lattices correspond to three ways in which the four lattice sites of
  a tetrahedron can be split into two pairs. The corresponding cluster decomposition rules are also displayed. In the figure, solid and empty circles denote $\sigma^z=1$ and $\sigma^z=-1$ respectively.}
\end{figure}

With this in mind, we define  three motifs, corresponding to the
three ways of decomposing four sites into two pairs of sites. These motifs specify the cluster decomposition rule for all plaquette operators on a given spatial plaquette in
the following way: For vertices with weight $8J$, all motifs do not yield valid microcanonical 
cluster decomposition rules if the pairing given by the motif is used to decompose
the eight legs into two a priori different clusters of four legs each. Only two out of the three motifs specify microcanonical cluster decompositions for a given spin configuration. The third motif implies a decomposition that, if used to perform a cluster update, can produce
vertices with zero weight which never occur in the operator string.
In this last case, when the motif is incompatible with a microcanonical cluster decomposition for a given spin configuration, all eight legs of such a vertex are assigned to a single cluster right at the outset. When the cluster decomposition implied by the motif results is a valid microcanonical
decomposition, the eight legs are split into two a priori different clusters of four legs each following this decomposition.
For the vertices with weight $6J$, every motif gives a valid cluster decomposition rule, which is followed. These motifs and the 
corresponding cluster decomposition rules are depicted in Fig.~\ref{pyr_rules}.

\begin{figure}[t]
  \includegraphics[width=8.5cm]{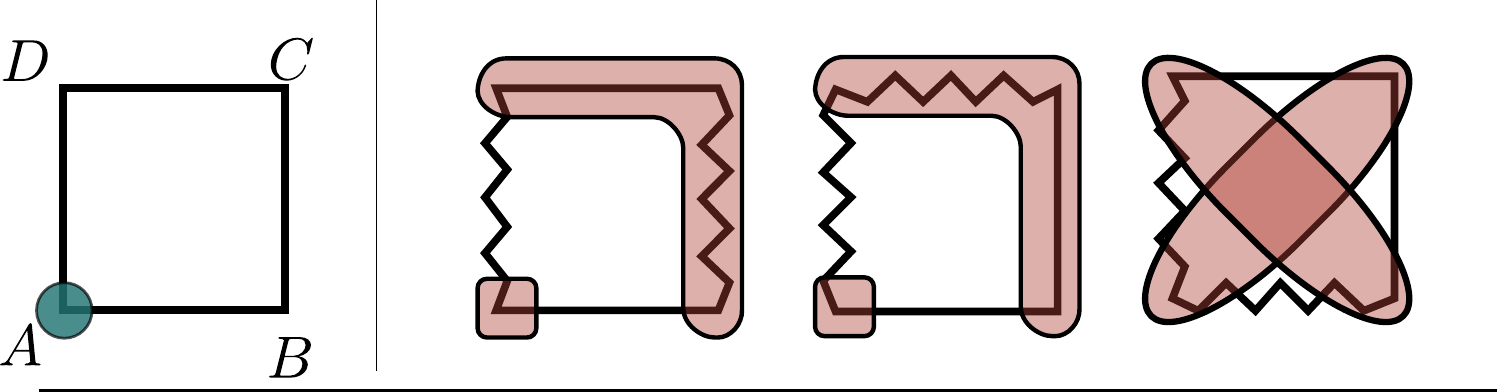}
  \caption{\label{egt_rules} The motifs for the TFIM with multi-spin interactions on the square lattice consist of a single privileged site. There are thus four
  possible motifs, corresponding to the four sublattices of sites of the square lattice.
  The cluster decomposition rule for one of these motifs is shown in the figure. The rule for the other motifs follows by analogy. In the figure, frustrated bonds are denoted by zigzag lines, while unfrustrated bonds are denoted by straight lines. }
\end{figure}
\subsubsection{TFIM with multi-spin interactions on the square lattice.}
For the TFIM with multi-spin interactions, decomposing the Ising-exchange Hamiltonian into
square plaquettes is anyway a more natural decomposition:
\begin{align}
    H_{\mathrm{TFIM}}=&-\sum_{i}H_{\tetr,p} -\sum_{i}H_{\mathbf{1},i} -\sum_{i}H_{\sigma^{x},i}\\
  \nonumber   H_{\Box,p}=&3J-J(\sigma^{z}_{A}\sigma^{z}_{B}\sigma^{z}_{C}\sigma^{z}_{D}  \\
  \nonumber &+\sigma^{z}_{A}\sigma^{z}_{C}  +\sigma^{z}_{B}\sigma^{z}_{D} )_p.
  \label{egt_decomp}
\end{align}
Here $p$ labels the square plaquettes of the lattice, and $A, B, C$ and $D$ denotes
the four sites making up the plaquette, labeled according to the four sublattice decomposition shown in Fig.~\ref{fftfim_lat}. All configurations with two frustrated bonds (antialigned pair
of nearest neighbour spins) corresponds to eigenvalue
$4J$ for $H_{\tetr,p}$, while configurations with four or zero frustrated bonds have eigenvalue $0$ and the corresponding plaquette operators do not appear in the operator string.

Our choice of motif in this case consists  of a single privileged site among the four that
make up a spatial plaquette. Thus each spatial plaquette has four possible motifs.
The motif on a given spatial plaquette determines the cluster decomposition of
plaquette operators at that location in the following way: If \emph{only one} frustrated bond touches the privileged site, the two legs corresponding to this site are 
assigned to a a priori different cluster, and the other six legs make up the other cluster (as usual, the two clusters could in principle merge at a future step in the cluster construction). If the privileged site is touched by two or zero frustrated bonds, then the four legs corresponding to the privileged site and its diagonally opposite site are assigned to one cluster,
and the other four legs to an a priori different cluster (which could in principle  merge with the other cluster at a future step in the cluster construction). These motifs and the cluster decomposition rules are illustrated in Fig.~\ref{egt_rules}.

\subsection{Premarking Strategies}
\label{premarkingstrategies}

With the motifs for each of the four models in place, one still has the freedom of choosing a motif for each spatial plaquette. As we have mentioned in the previous section, there are two such choices on the square lattice for the fully frustrated TFIM and four such choices for the TFIM with multi-spin interactions, while there are three such choices for the TFIM on the pyrochlore and planar pyrochlore lattices. Here we specify the premarking strategies used
to imprint one of these motifs on each spatial plaquette in each of these systems.

\subsubsection{Configuration-independent premarking : Swendsen-Wang updates}
The simplest way to implement premarking is to premark a motif for each spatial plaquette before each cluster update is initiated. Then, the  cluster construction procedure assigns the legs on \emph{all the Ising exchange operators} to different clusters following the prescription implied by the premarked motif on the spatial location of the operator. Using this prescription, the entire SSE operator string can be decomposed into non-overlapping clusters using the union-find algorithm. Each of these clusters is then flipped with a probability of $1/2$, following the approach of Swendsen and Wang.\cite{Swendsen_Wang}

The choice of motif on each spatial plaquette leads to a wide variety of Swendsen Wang updates, and a premarking pattern can be selected based on the problem at hand. For the fully frustrated TFIM  on the square lattice, we find the best premarking choices to be 
either premarking all plaquettes with the motif that picks A and C sublattice sites, or premarking all plaquettes with the motif that picks B and D sublattice sites. This yields two different
Swendsen Wang update schemes, which we deploy independently during our Monte Carlo run (either
by picking any one of them at random at each Monte Carlo step (MC step) or by alternating between them).

For the TFIM with multi-spin interactions on the
square lattice, we find it useful to set the privileged site of all spatial plaquettes to
be either the $A$, or the $B$ or the $C$ or the $D$ sublattice site on that plaquette.
This gives four possible Swendsen Wang update schemes which we deploy independently during
our Monte Carlo run (either by picking any one of them at random at each Monte Carlo
step, or by cycling through them).

For the TFIM on the planar pyrochlore lattice, we have obtained
the best results by randomly and independently choosing between the motifs on the first and third rows of Fig.~\ref{pyr_rules} for each spatial plaquette at the beginning of each Swendsen Wang cluster update (using the motif in the second row of Fig.~\ref{pyr_rules}  leads to slow Monte Carlo dynamics when the system has a tendency to Neel order at low temperature). For the TFIM on the pyrochlore lattice, we have
randomly and independently chosen one of the three motifs  in Fig.~\ref{pyr_rules} for each spatial plaquette at the start of the Swendsen Wang cluster update.

\subsubsection{Dynamic premarking : Wolff Updates}
Now, we present a different premarking strategy that yields an efficient Wolff cluster update
procedure that can dramatically outperform the Swendsen Wang cluster updates
for the planar pyrochlore and pyrochlore lattices in some regimes (we have also explored this in the other systems studied here, and find some improvement, but confine ourselves here to describing the procedure for the pyrochlore lattices). 
In this approach, a random leg is chosen to start growing a cluster. After the cluster is grown, all legs (spin-states) 
belonging to the cluster are flipped with probability one. One Monte Carlo step is
then defined as a certain fixed number of Wolff cluster updates alternating with diagonal
updates.

As described in the previous section, the two kinds of Ising-exchange vertices in the pyrochlore lattices have weights $6J$ and $8J$. Whenever an Ising-exchange operator living
on a particular spatial plaquette is first encountered (and this can be either an operator
with weight $6J$ or with weight $8J$), a motif is chosen for that spatial plaquette and
marked on it. This motif is chosen in a manner that provides the most optimal decomposition of the legs of this first
Ising-exchange operator encountered at that location: More precisely, if the vertex first
encountered during cluster construction has weight $6J$, then one of three motifs described in  Fig.~\ref{pyr_rules} is chosen with equal probability and imprinted on that spatial plaquette.

On the other hand, if the vertex first encountered is an $8J$ vertex, then we randomly
choose one of the two motifs that gives a consistent microcanonical decomposition of
the legs of that vertex (into two groups of four legs each, assigned to two
a priori different clusters). The third motif, which  assigns all legs of the vertex to a single cluster, is {\em never} chosen at this first encounter. The motif thus chosen is then imprinted on the
spatial plaquette. In this manner, the motif on each spatial plaquette gets assigned
when the cluster construction first encounters an Ising exchange operator living on that
plaquette. All subsequent Ising exchange operators living on the same spatial
plaquette are then decomposed into clusters in accordance with the rules specified
by the motif that has already been imprinted on that spatial plaquette. When one cluster
is constructed in this way, it is flipped with probability one. A Monte Carlo step is
defined as a certain fixed number of such Wolff cluster flips interspersed with diagonal
updates.

As we demonstrate below, this procedure leads to further improvement in the cluster size
statistics, and consequently leads to vastly improved auto-correlation times in some regimes. While the implementation of this Wolff type procedure is more involved, and the CPU time required in our implementations  are typically larger than an equivalent number of
Swendsen Wang updates, the Monte Carlo autocorrelation times can be significantly
smaller for some observables.

\section{ One-dimensional clusters: Canonical update}
\label{canonical}
\begin{figure}[t]
  \includegraphics[width=5cm]{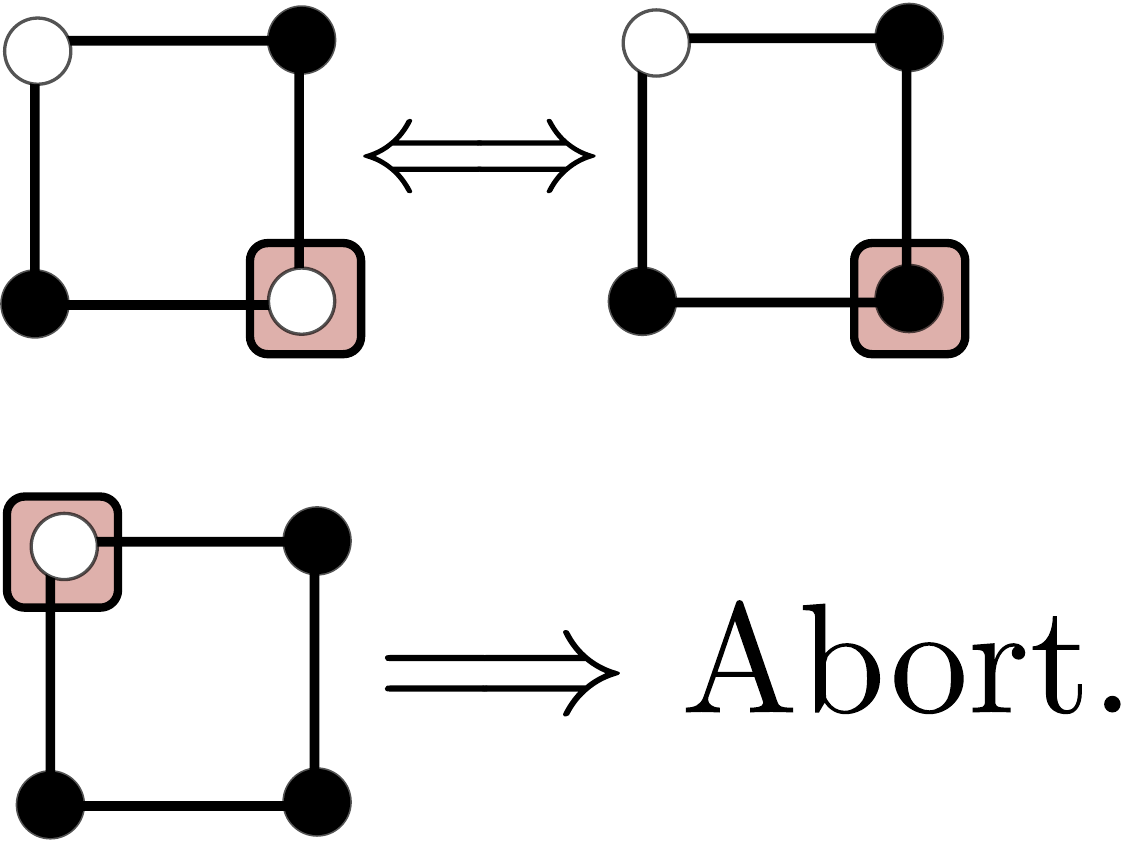}
  \caption{\label{pyrcan_rules} The canonical update is aborted only if it encounters
    a minority leg of an Ising-exchange vertex of the transverse field Ising model on the pyrochlore or planar pyrochlore lattice.}
\end{figure}
For the pyrochlore lattices, the Ising-exchange vertices appear with two different weights: Vertices with weight $6J$ correspond to three spins on a tetrahedron  being aligned together, and the fourth
antialigned with these, while vertices with weight $8J$ correspond to ice-rule states in which
two spins point up and the other two point down. The microcanonical cluster updates described in the previous sections keep the weight of each vertex unchanged. Here we demonstrate the utility of a simple canonical update scheme, which acts at only one spatial
site of the lattice, but has the virtue of being able to switch between operators of
different weight acting at that spatial site. This procedure grows ``one dimensional clusters'' in the ``imaginary time'' direction (Indeed, we label the canonical update as '1D' in all the figures), which are the QMC analog of single-spin flip Metropolis updates in the classical case.
We also introduce such canonical updates for the other TFIMs considered in Sec.~\ref{models}. Note however that the microcanonical cluster update described earlier also makes one-dimensional clusters some fraction of the time in these other cases, and the improvement due to the canonical one-dimensional update is therefore expected to be less significant in these cases.

\subsection{TFIM on the pyrochlore lattices.}
The canonical update procedure for the TFIM on the pyrochlore lattice is specified by the following steps:
\begin{enumerate}
  \item Choose a random leg $l_0$ belonging to a randomly chosen single site operator ($H_{\mathbf{1},i}$ or $H_{\sigma^{x},i}$), and set $l_i=l_0$
  \item Assign $l_i$ to the one dimensional cluster, and  also assign the leg $l_j$ which is linked to $l_i$ to the cluster.
  \item If $l_j$ belongs to a single-site vertex, terminate cluster construction. Go to 6.
  \item Else, if $l_j$ belongs to vertex with weight $8J$, then set $l_i=l_k$ and go to 2.
    Here, $l_k$ is the other leg of the same vertex which also belongs to the same site as $l_j$.
   \item Else, if $l_j$ belongs to a vertex with weight $6J$, then the action depends on the spin state at $l_j$:
     \begin{itemize}
     \item  If, $l_j$ has a majority spin on it , then set  $l_i=l_k$. Go to 2.
       Here $l_k$ is the other leg of the same vertex, which belongs to the same site as $l_j$.
       \item If $l_j$ has a minority spin on it, then abort the cluster construction. Go to 7.
 \end{itemize}
 \item Flip the spin-states corresponding to all legs in the cluster. End.
 \item Abort. Empty the cluster. Start a new update without flipping any spin-states.
\end{enumerate}

This update enables fluctuations between vertices of weight $6J$ and $8J$. As described above, the one dimensional cluster-construction
procedure aborts when it encounters a minority spin-state at a $6J$ vertex. The cluster construction rule is illustrated in
Fig.~\ref{pyrcan_rules}.

\begin{figure}[t]
  \includegraphics[width=5cm]{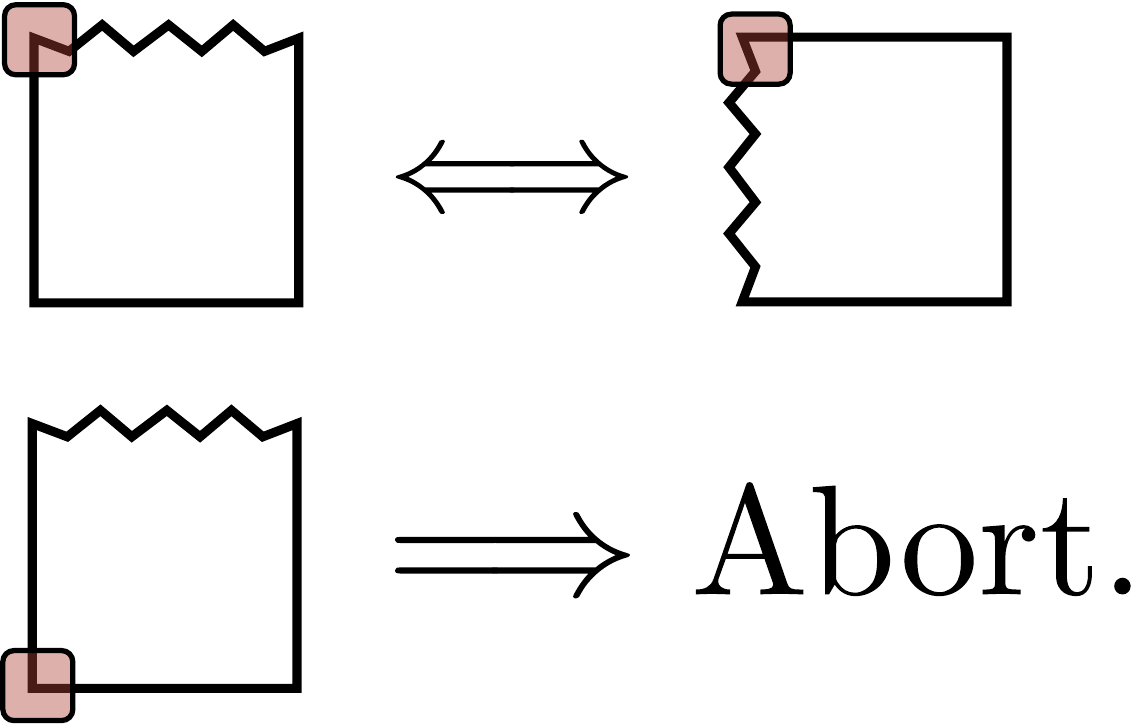}
  \caption{\label{fftfimcan_rules} The canonical cluster construction is aborted only
  if it would result in a vertex with zero weight for the FFTFIM.}
\end{figure}
\subsection{FFTFIM on the square lattice.}
For the FFTFIM on the square lattice, all Ising exchange vertices have the same weight. The one-dimensional updates are thus not ``canonical'' in this case.
The one dimensional cluster construction rules 
are specified by the following steps :
\begin{enumerate}
  \item Choose a random leg $l_0$ belonging to a randomly chosen single site operator ($H_{\mathbf{1},i}$ or $H_{\sigma^{x},i}$), and set $l_i=l_0$
  \item Assign $l_i$ to the one dimensional cluster, and  assign the leg $l_j$ which is linked to $l_i$ (in the operator string) to this cluster.
  \item If $l_j$ belongs to a single-site vertex, terminate cluster construction. Go to 5.
   \item Else,   the action depends on the spin states at this Ising exchange vertex, which
    lives on a plaquette with one frustrated bond.
     \begin{itemize}
     \item  If, $l_j$ belongs to a spatial site which is touched by the frustrated bond, then set ta set $l_i=l_k$. Go to 2. Here the leg $l_k$ is the other leg of this vertex which belongs
       to the same spatial site as $l_j$.
       \item Otherwise, abort the cluster construction. Go to 6.
 \end{itemize}
 \item  Flip the spin-states corresponding to all legs in the cluster. End.
 \item Abort. Empty the cluster. Start a new update without flipping any spin states.
\end{enumerate}
This ``canonical'' update for the FFTFIM on the square lattice is illustrated in Fig.~\ref{fftfimcan_rules}.

\subsection{TFIM with multi-spin interactions.}
\begin{figure}[t]
  \includegraphics[width=5cm]{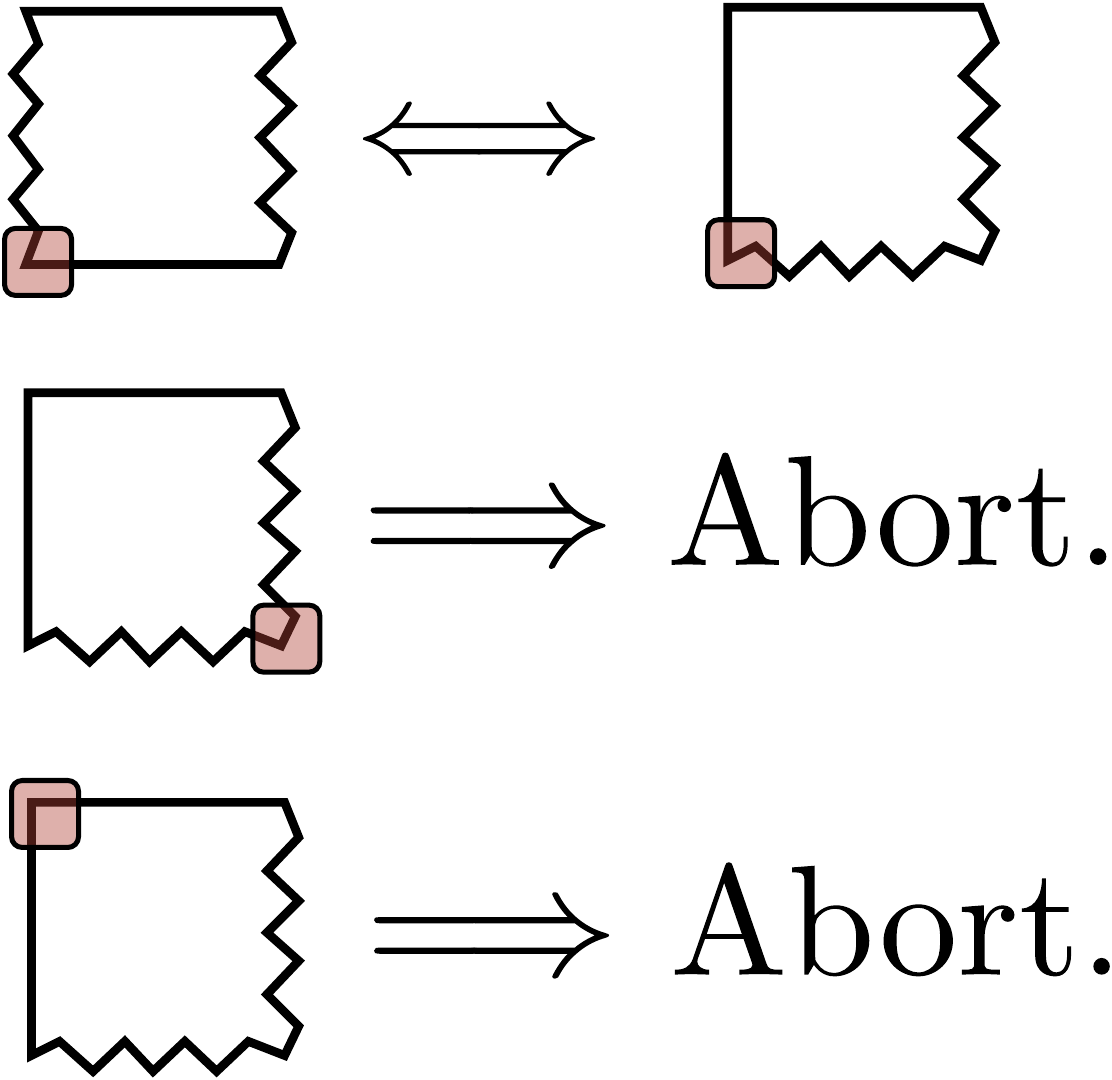}
  \caption{\label{egtcan_rules} The canonical cluster construction is aborted only if it would
    result in a vertex with zero weight for the TFIM with multi-spin interactions on the
    square lattice.}
\end{figure}

Finally, we describe the canonical one-dimensional cluster update for  the square lattice TFIM with multi-spin interactions, introduced in Sec.~\ref{models}.
The cluster construction rules are specified as follows :
\begin{enumerate}
  \item Choose a random leg $l_0$ belonging to a randomly chosen single site operator ($H_{\mathbf{1},i}$ or $H_{\sigma^{x},i}$), and set $l_i=l_0$
  \item Assign $l_i$ to the one dimensional cluster, and  assign the leg $l_j$ which is linked to $l_i$ (in the operator string) to the cluster.
  \item If $l_j$ belongs to a single-site vertex, terminate cluster construction. Go to 5.
  \item Else,   the action depends on the spin states on the Ising exchange vertex to which
    $l_j$ belongs. This vertex lives on a spatial plaquette in which there are two pairs
    of nearest neighbour antiparallel spins, which we identify as frustrated bonds.
     \begin{itemize}
     \item  If, $l_j$ belongs to a spatial site which is touched by one frustrated bond, then set $l_i=l_k$. Go to 2. Here $l_k$ is the other leg of the same vertex which belongs to the
       same spatial site as $l_j$.
       \item Otherwise, abort the cluster construction. Go to 6.
 \end{itemize}
 \item  Flip the spin-states corresponding to all legs in the cluster. End.
 \item Abort. Empty the cluster. Start a new update without flipping any spin-states.
\end{enumerate}
This cluster construction rule is illustrated in Fig.~\ref{egtcan_rules}.

\section{Performance}
\label{Performance}
To test our implementations of the new cluster algorithm (including the canonical
cluster update), we have compared high-precision results for small sizes with
the results of exact diagonalization 
and checked that they agree within statistical errors. 

To test the efficiency of our algorithms, we simulate finite size systems with periodic boundary conditions using both the new plaquette percolation based cluster updates (with and
without the canonical updates), 
and the conventional link percolation based
cluster updates described in Sec.~\ref{review}. In all our simulations, we monitor
Monte Carlo autocorrelation functions of observables relevant to the model in question.
For an observable $O$, with Monte Carlo estimator given by $O_{\mathrm{MC}}(n)$ in configuration $n$, the autocorrelation function  $A_{O}(\tau_{\mathrm{MC}})$  is defined as
\begin{equation}
  A_{O}(\tau_{\mathrm{MC}})=\frac{\langle O_{\mathrm{MC}}(n+\tau_{\mathrm{MC}})O_{\mathrm{MC}}(n)\rangle -\langle O_{\mathrm{MC}}(n)\rangle ^{2}}{\langle [O^2]_{\mathrm{MC}}(n) \rangle-\langle O_{\mathrm{MC}}(n)\rangle ^2} \; ,
  \label{autocorreq}
\end{equation}
where the angular brackets denote average of the estimator $O_{\mathrm{MC}}(n)$ over the configurations $n$ generated in the Monte Carlo simulation.
For later reference, we also introduce $\chi_{O}$ the static susceptibility corresponding to a local operator $O$, as 
\begin{equation}
\chi_{O}=\frac{L^{2}}{\beta}\langle\lvert \int_{0}^{\beta} d\tau  O(\tau)\rvert^{2}\rangle,
\end{equation}
where $O(\tau)$ is the imaginary-time Heisenberg operator corresponding to the Schrodinger operator $O$, given by $O(\tau)=\exp(H(\tau))O \exp(-H(\tau))$ and the angular bracket is over the canonical ensemble at inverse temperature $\beta$.
Analogous to Eq.~\eqref{autocorreq}, we we also definte the autocorrelation function of the static susceptibility $\chi_O$ as $A^{\chi}_{O}$.

\begin{figure}[t]
  \includegraphics[width=\columnwidth]{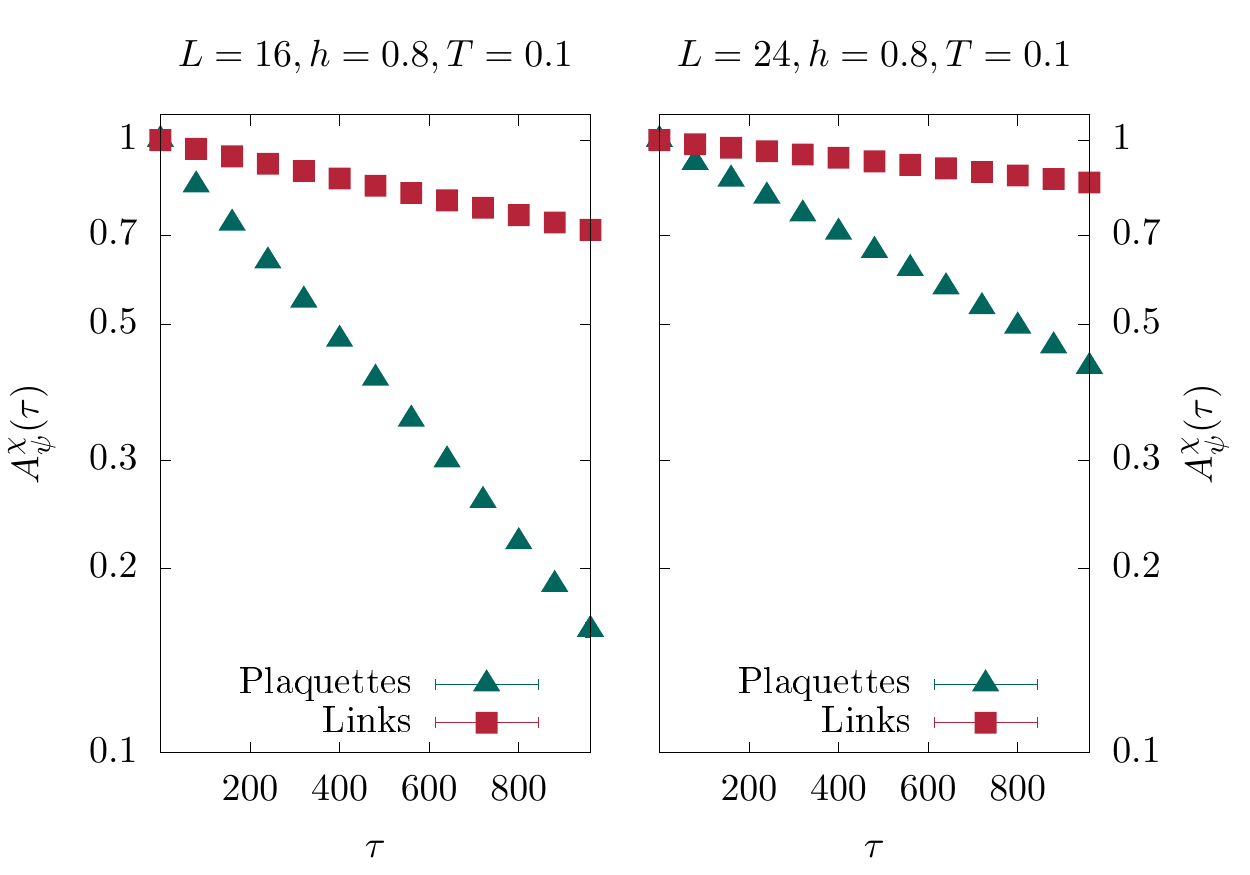}
  \caption{\label{fftfim_autos} Comparison of autocorrelation function for the Monte Carlo estimator of $\chi_{\psi}$ (defined in text)  in the FFTFIM on a $L \times L$ square lattice using both the link percolation based cluster algorithm and the plaquette percolation based cluster algorithm.}
\end{figure}
\begin{figure}[t]
  \includegraphics[width=\columnwidth]{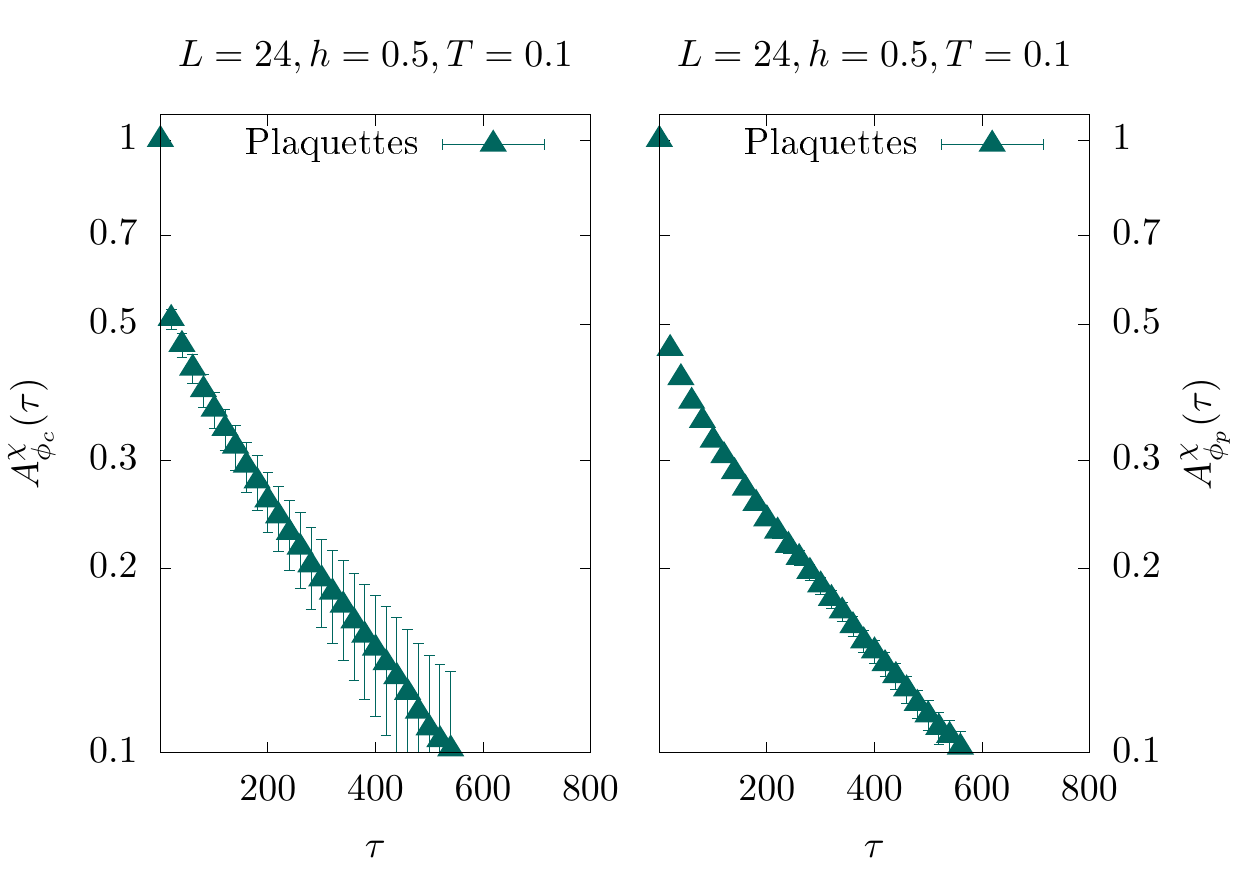}
  \caption{\label{egtautos}Comparison of autocorrelation function for the Monte Carlo estimator of $\chi_{\phi_c}$ and  $\chi_{\phi_p}$ (defined in text)  in the TFIM with
    multispin interactions on a $L \times L$ square lattice.}
\end{figure}

\subsubsection{Fully frustrated TFIM on the square lattice.}
For the FFTFIM on the square lattice, the columnar and plaquette states (using the terminology of the dual dimer model that emerges in the limit of $T \lesssim \Gamma \ll J$) are both described in terms of a nonzero expectation value for the complex order parameter
$\psi$:
\begin{align}
\nonumber  \psi =& \frac{1}{L^2} [\sigma_A^{z}\exp(i 3\pi/8) +\sigma_B^{z}exp(i \pi/8) \\
  &+\sigma_C^{z}\exp(-i \pi/8) +\sigma_D^{z}\exp(-i3\pi/8) ].
  \label{fftfimoparam}
\end{align}
$\sigma^{z}_{\alpha}$ refers to the sum of all spins on the corresponding sublattice $\alpha$ (as defined by the four sublattice decomposition of the square lattice shown in Fig.~\ref{fftfim_lat}).

We have computed  the static susceptibility $\chi_{\psi}$  corresponding to this order parameter, as well as its Monte Carlo autocorrelation function $A^{\chi}_{\psi}$. For the plaquette percolation based scheme, we have defined one Monte Carlo step to be
two Swendsen-Wang type cluster updates, each of which immediately follow a diagonal update. In the first cluster update of any Monte Carlo step, we premark, when
using the plaquette percolation approach, all sites belonging to sublattices A and C, while for the  second cycle, we premark all sites belonging to sublattices B and D.
For direct comparison with the conventional link percolation based cluster algorithm, we have similarly defined one Monte Carlo step of that algorithm to again be two cluster
updates with a diagonal update immediately preceding each cluster update.
For the autocorrelation measurements, we have used a warm-up of $2\times10^5$ MC steps, followed by a run of $2\times10^6$ MC steps.
We display the autocorrelations of the order parameter $A^{\chi}_{\psi}$ in Fig.~\ref{fftfim_autos}. Clearly, the algorithm based on plaquette percolation is much more 
efficient in decorrelating Monte Carlo configurations from each other.

\begin{figure}[t]
  \includegraphics[width=\columnwidth]{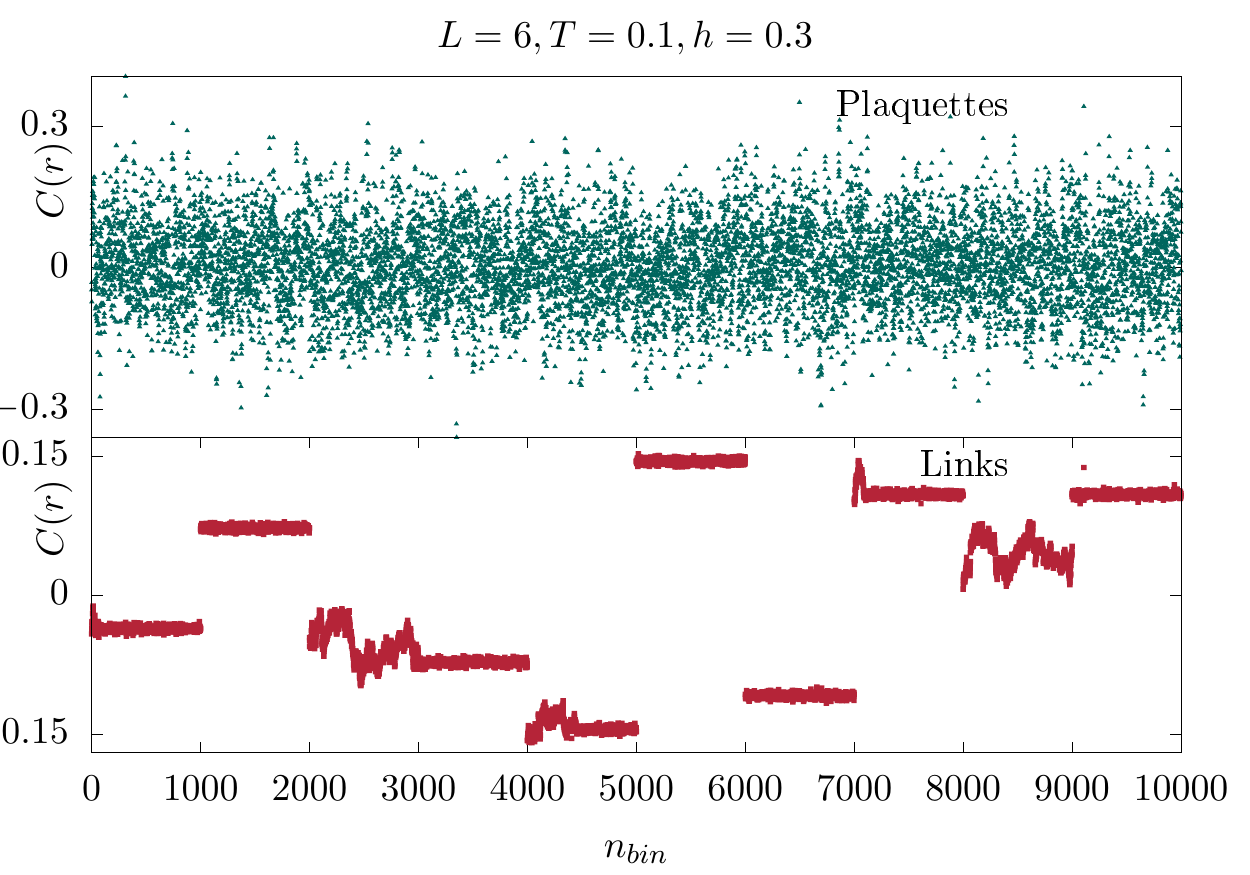}
  \caption{\label{sztimes}Time series of Monte Carlo estimator of $C$, the equal time correlation function of $\sigma^z$, recorded during the simulation of the TFIM on a $L \times L \times L$ pyrochlore lattice, obtained by the link percolation based cluster algorithm and the plaquette percolation based cluster algorithm.}
\end{figure}
\subsubsection{TFIM with multi-spin interactions on the square lattice.}
As described in Sec.~\ref{models}, there are two natural ordered states for the TFIM with multi-spin interactions.
The columnar state can be described by a complex order parameter, $\phi_c$, given by
\begin{equation}
\phi_c= \frac{1}{L^2} \sum_{\vec{r}} \left [(-1)^x \sigma^z(\vec{r})+i (-1)^y\sigma^z(\vec{r}) \right ].
\end{equation}
The other natural state is the plaquette phase, which can be described by the following
off-diagonal order parameter
\begin{equation}
  \phi_p=\frac{1}{L^2} \sum_{\vec{r}} \left[ (-1)^{x+y}\sigma^{x}(\vec{r}) \right ].
\end{equation}
For the TFIM with multi-spin interactions, there is no link-based decomposition of the Hamiltonian,
and hence no conventional cluster algorithm. To test the efficiency of the plaquette percolation based algorithm, we compute the static susceptibilities 
$\chi_{\phi_c}$ and $\chi_{\phi_p}$, and study the corresponding Monte Carlo autocorrelation functions  $A^{\chi}_{\phi_c}$ and $A^{\chi}_{\phi_p}$
. One MC step is defined to consist of four diagonal updates, each followed immediately by a cluster update. In these four cluster updates we
cycle through the four possible motifs in turn.
For the autocorrelation measurements, we have performed simulations with a warm-up of $2\times10^5$ MC steps, followed by a run of $2\times10^6$ MC steps. 
These autocorrelations are displayed in Fig.~\ref{egtautos}, and it is clear that
the algorithm provides a fairly efficient means of simulating the physics of this system.

\begin{figure}[t]
  \includegraphics[width=\columnwidth]{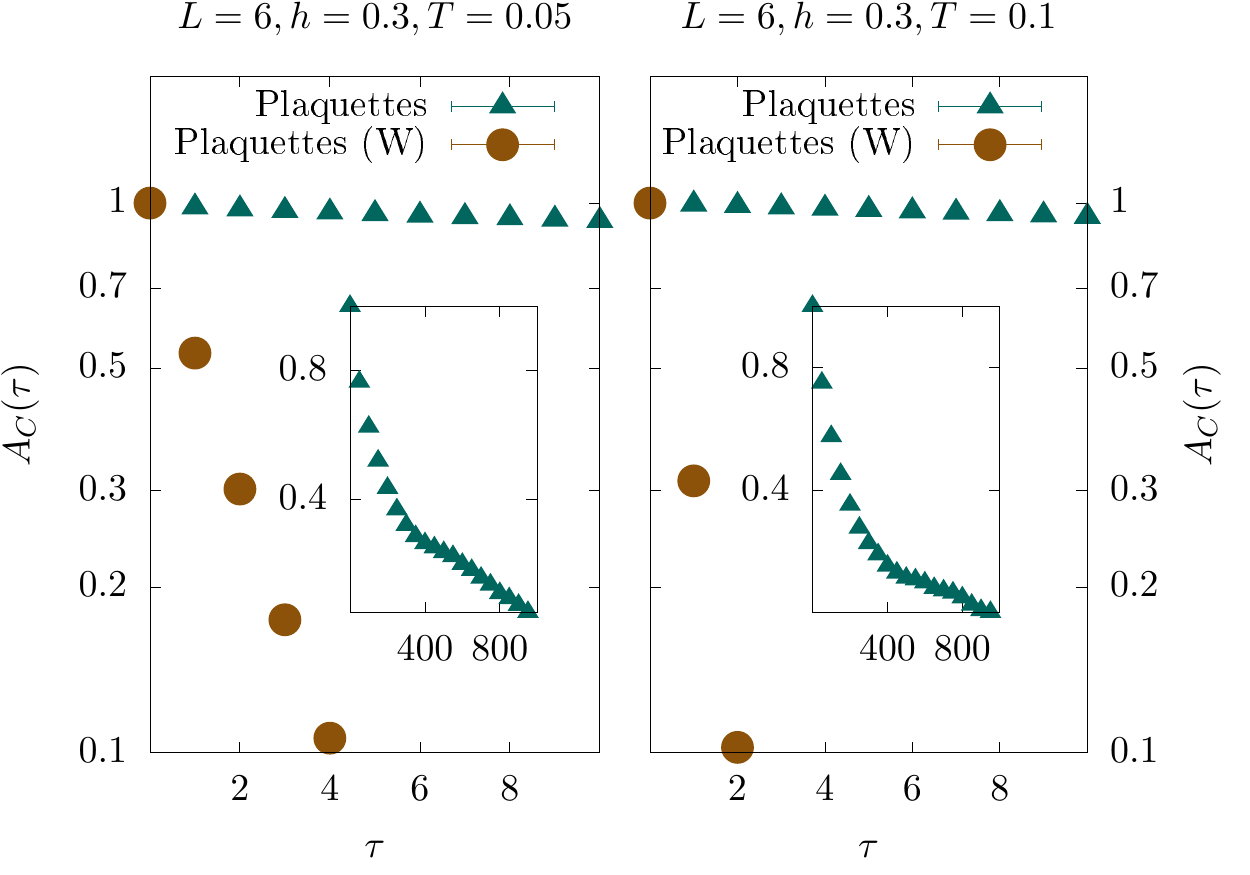}
  \caption{\label{pyrautos}Autocorrelation function for the Monte Carlo estimator of the equal time correlation function $C$ of the TFIM on a  $L \times L \times L$ pyrochlore lattice for the plaquette percolation based cluster algorithm using both the Swendsen-Wang updates with configuration-independent premarking (labelled `Plaquettes') and the Wolff updates with dynamic premarking (labelled `Plaquettes(W)') described in Sec.~\ref{premarkingstrategies}  }
\end{figure}
\subsubsection{TFIM on the pyrochlore lattice.}
The TFIM on the pyrochlore lattice has no long-range order, and is expected to host an interesting Coulomb  liquid 
ground state. In the absence of any natural order parameter to look at, we focus on the correlations
of $\sigma_z$. To compare algorithms, we first look at the correlation function $C$ for a finite size 
system of $L\times L$ unit cells with periodic boundary conditions. $C$ is defined
as
\begin{equation}
  C= \frac{1}{L^2}\sum_{x,y,z}\sigma^z(x,y,z)\sigma^z(x+\frac{L}{2}, y+\frac{L}{2}, z+\frac{L}{2}) \; .
  \label{corrfnsz}
\end{equation}
In addition, we measure the mean value of $\sigma^x$ averaged over the entire sample.

To enable direct comparison of autocorrelation functions, we define a Monte Carlo step for each of the algorithms to involve the same amount of work as follows
For both the conventional algorithm as well as the new algorithm, we define one MC step as two diagonal updates, each followed immediately by a Swendsen-Wang type cluster update. For the TFIM on the pyrochlore lattice, we also consider the Wolff-type version of the plaquette percolation algorithm, introduced in Sec.~\ref{premarkingstrategies}. 
In this case, we define one MC step to be two diagonal updates, each followed
immediately by constructing $n_{\mathrm{wolff}}$ Wolff-type clusters and flipping them
with probability one.
$n_{\mathrm{wolff}}$ is set to be the integer part of the average of $n_{\mathrm{tot}}/{n_c}$ (measured during the warmup), where $n_{\mathrm{tot}}$ is the total number of legs in the 
linked-list and $n_c$ is the number of legs in a Wolff-type cluster. For each method,
we have performed simulations
with a warm-up of $2\times10^5$ MC steps, followed by a run of another $2\times10^5$ steps.

\begin{figure}[t]
  \includegraphics[width=\columnwidth]{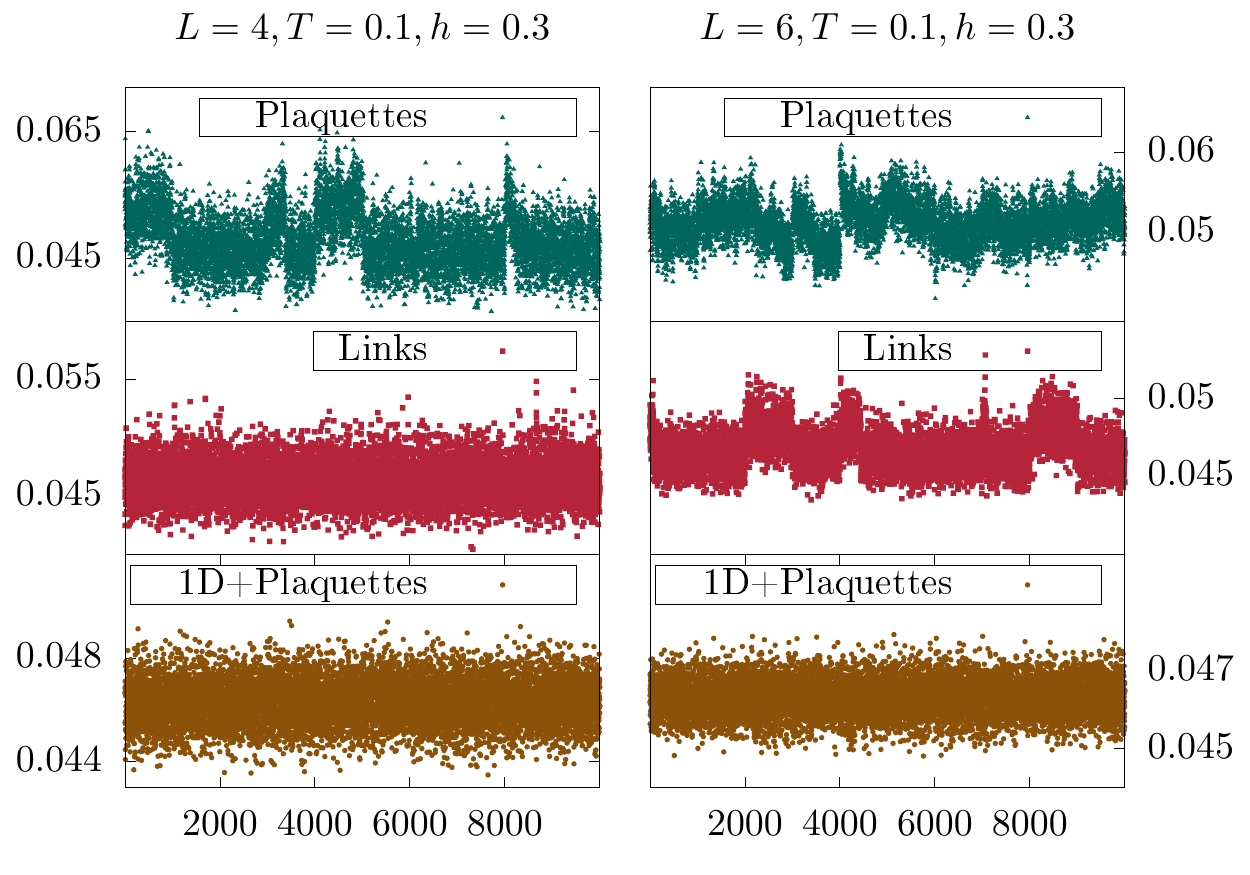}
  \caption{\label{sxtimes}Time series of Monte Carlo estimator of the mean value of
    $\sigma^x$ recorded during the simulation of the TFIM on a $L \times L \times L$ pyrochlore lattice by the link percolation based cluster algorithm and the plaquette percolation based cluster algorithm with and without the canonical (`1D') updates.  }
\end{figure}

\begin{figure}[t]
  \includegraphics[width=\columnwidth]{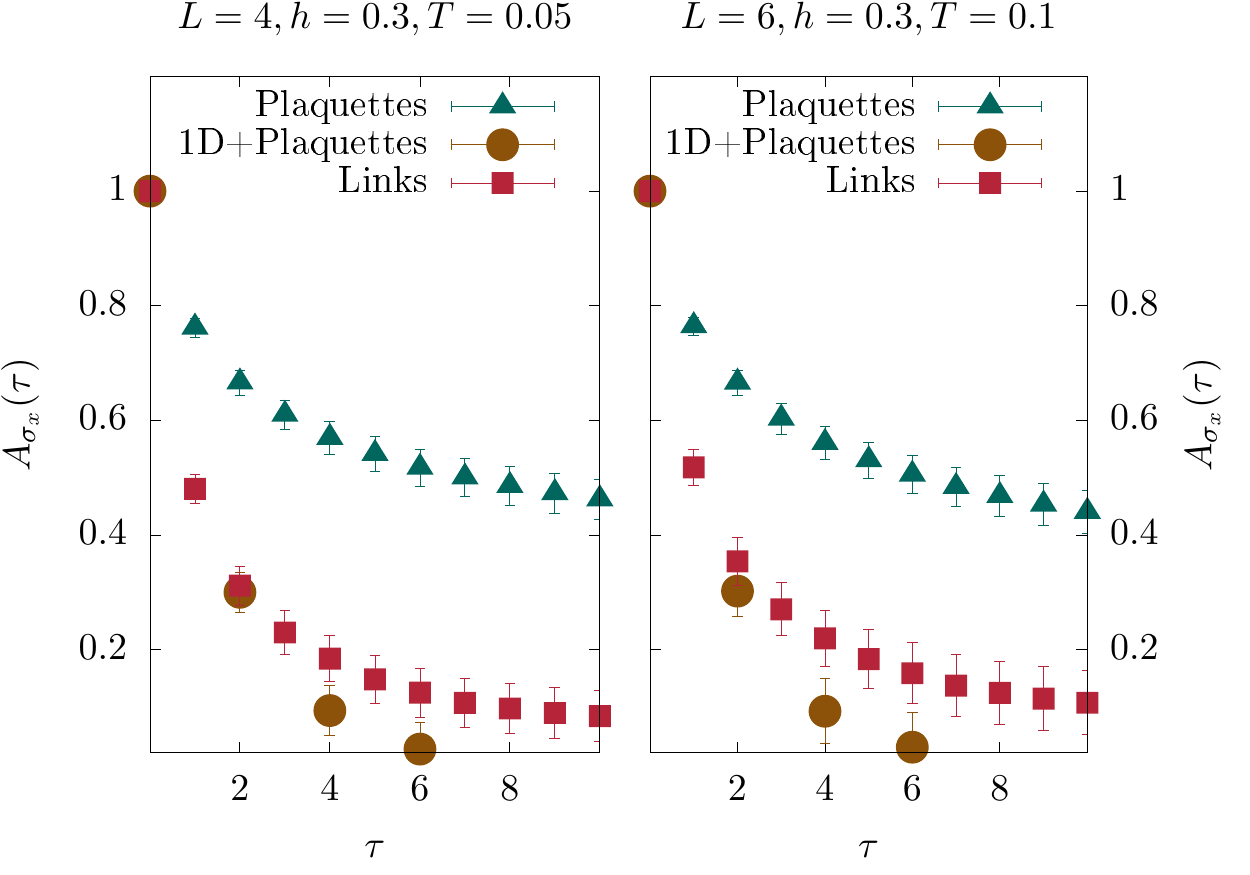}
  \caption{\label{sxpyrautos} Autocorrelation function for the Monte Carlo estimator of $\langle \sigma_x \rangle$ in the TFIM on a  $L \times L \times L$ planar pyrochlore lattice for the link percolation based cluster algorithm and the plaquette percolation based cluster algorithm with and without the canonical (`1D') cluster updates.}
\end{figure}

Several things are apparent from the results shown in Figs.~\ref{sztimes},~\ref{pyrautos},~\ref{sxtimes},~\ref{sxpyrautos}.  First, we note that the conventional link percolation based algorithm suffers from severe loss of ergodicity, as is evident from the time series for the
bin averages (over bins made up of $100$ consecutive MC steps) of the estimator of the equal
time correlation function $C$ of $\sigma^z$ at distance $L/2$. As a result, the Monte Carlo result for $C$ simply does not converge. This is clear from plots of this time series for $10$ independent runs corresponding to different
initial configurations (Fig.~\ref{sztimes}). Clearly, the bin averages for a particular seed tend to 
stick around values which are strongly dependent of the initial conditions, showing lack of ergodicity. 
We also show, in Fig.~\ref{sztimes}, that there is no such loss of ergodicity for the plaquette percolation based
algorithm. Bin averages for different seeds are indistinguishable from each other, indicating
that the estimator of $C$ is accurate.
We have not calculated the corresponding autocorrelation function $A_{C}$ for the conventional link percolation based algorithm,
as $C$ does not converge. However, we display the autocorrelation function $A_{C}$ for the plaquette percolation based algorithm
with configuration-independent premarking and Swendsen-Wang updates, and compare this with the alternate implementation that uses Wolff-type clusters and a dynamic premarking
scheme (Fig.~\ref{pyrautos}) for a more challenging parameter regime compared to the display of  bin averages in Fig.~\ref{sztimes}. We find that dynamic premarking leads to further, and quite obvious improvements in autocorrelation times in this challenging regime.

Now, we shift our focus to the mean value of $\sigma^{x}$. In Fig.~\ref{sxtimes}, we display the bin-averages for $\sigma^{x}$. We see that 
even for relatively modest sizes, the plaquette percolation based algorithm has some
difficulty fully equilibrating the estimators of $\sigma^x$, while the link percolation
based algorithm seems to fare somewhat better for smaller sizes. 
We have found this somewhat poor equilibration of $\sigma^x$ in plaquette-percolation based approaches to be
a common feature of both the pyrochlore lattice as well as the planar pyrochlore lattice. As mentioned earlier, this is due to the fact that the plaquette-based quantum cluster algorithms do not grow one-dimensional single-spin clusters if the minimum exchange-energy configurations are charcterized by a constraint on the net spin of individual plaquettes. As noted earlier, such one-dimensional clusters can be grown by a canonical procedure that allows fluctuations between Ising exchange operators of different weights ($6J$ of $8J$) in the plaquette-based formulation of QMC. We define one MC step of this modified algorithm as being two cycles of a diagonal update, followed immediately by a Swendsen-Wang type cluster update with configuration-independent premarking (or $n_{\mathrm{wolff}}$ Wolff type clusters
grown with dynamic premarking), followed by $n_{1d}$ canonical updates. Just like $n_{\mathrm{wolff}}$,  $n_{1d}$ is set to the average of $n_{\mathrm{tot}}/n_c$, measured during the warm-up. $n_{\mathrm{tot}}$  is the total number of legs in the 
linked-list as before, while $n_c$ is now the number of legs in each canonical cluster.

We find that the plaquette percolation based algorithm 
supplemented with these canonical updates equilibrates $\sigma^x$ fairly well, even for larger sizes, where the conventional link percolation based
algorithm develops some equilibration problems (Fig.~\ref{sxtimes}) at these larger sizes. Recent work\cite{Emont_Wessel} has also explored a ``greedy'' implementation of the triangular-decomposition based approach of Ref.~\onlinecite{Biswas_Rakala_Damle}, {\em i.e.} an approach in which one dispenses with the notion of premarked privileged sites imprinted on the {\em spatial} lattice, and allows single-spin flips of one of the (randomly chosen) majority spins of every triangular plaquette operator whenever these are encountered during the microcanonical cluster construction. As noted there, this ``greedy'' approach gives results comparable to the original link-based method, and is therefore expected to have similar equilibration problems at larger sizes.

Since these equilibration problems are not as severe as those for $C$, we have calculated the autocorrelation function $A_{\sigma^x}$ corresponding to $\sigma^x$
for different algorithms and displayed it in Fig.~\ref{sxpyrautos}. To compare the algorithm
that uses both plaquette percolation based microcanonical clusters and the canonical clusters with others that do not use the canonical clusters, we have 
scaled the $x$-axis of the data corresponding to the combined algorithm by a factor by 2; this is a conservative estimate of the 
computation time taken to do an extra set of cluster updates in every MC step in this combined approach.
We conclude that the canonical updates are crucial to
equilibrate $\sigma^{x}$ in the plaquette percolation based cluster update, and this combined
approach is clearly superior to the conventional link percolation based approach, and by
corollary (appealing to the results of Ref.~\onlinecite{Emont_Wessel}), superior to the ``greedy'' variant of the triangular-plaquette based approach explored recently.\cite{Emont_Wessel}

\subsubsection{TFIM on the planar pyrochlore lattice.}

\begin{figure}[t]
  \includegraphics[width=\columnwidth]{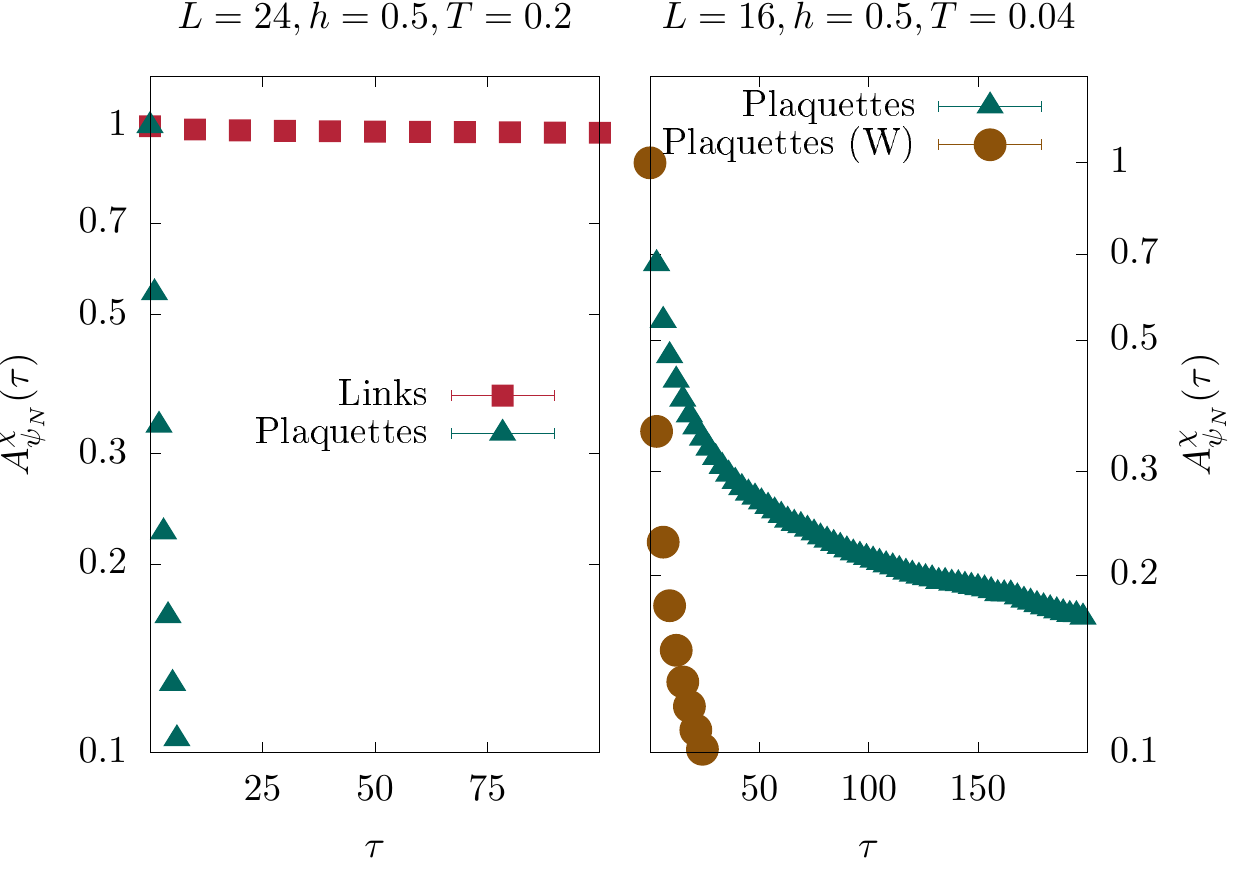}
  \caption{\label{planpyrautos}Autocorrelation function for the Monte Carlo estimator of $\chi_{\psi_{N}}$ (defined in the text) in the TFIM on a $L \times L$ planar pyrochlore lattice for the link percolation based cluster algorithm and the plaquette percolation based cluster algorithm using both the Swendsen-Wang updates with configuration-independent premarking (labelled `Plaquettes') and the Wolff updates with dynamic premarking (labelled `Plaquettes(W)') described in Sec.~\ref{premarkingstrategies}.}
\end{figure}
The TFIM on the planar pyrochlore lattice is known to be Neel ordered at low temperatures and not too low transverse magnetic fields.\cite{Henry_Roscilde}. 
The authors of Ref.~\onlinecite{Henry_Roscilde} 
also predict a plaquette-VBS phase at  extremely low temperatures and low magnetic fields. We do not expect such low temperatures
to be accessible to the initial simulations shown here by way of tests of our algorithm. Therefore, for testing purposes, we calculate  the Neel order parameter $\psi_N$, given by
\begin{equation}
  \psi_N=\sum_r (\sigma^z_{h(\vec{r})}-\sigma^z_{v(\vec{r})}) \; ,
\end{equation}
where $\vec{r}$ denotes sites of the dual lattice and $h(\vec{r})$ and $v(\vec{r})$
denote spin degrees of freedom located at $\vec{r}+\hat{e}_x/2$ and $\vec{r}+\hat{e}_y/2$, as  illustrated in Fig.~\ref{planpyr_lat}.
For the conventional algorithm based on link percolation, and for the plaquette percolation algorithm with
configuration-independent premarking, we define one MC step as two diagonal updates, each
followed immediately by a Swendsen-Wang type cluster update.
For the plaquette percolation based cluster algorithm with dynamic premarking, we replace the Swendsen-Wang type cluster update with $n_{\mathrm{wolff}}$ successive Wolff-type cluster updates.

For calculating autocorrelations, we perform 
simulations with a warm-up of $2\times10^5$ MC steps, followed by a run of $2\times10^6$ MC steps.
As in the three dimensional pyrochlore case, we also test a combined
approach in which microcanonical cluster updates are supplemented
by the canonical updates introduced in Sec.~\ref{canonical}. For this combined approach, we define one MC step to be two diagonal update, each followed immediately by a single Swendsen-Wang type cluster update (or $n_{\mathrm{wolff}}$ Wolff type cluster updates that use dynamic premarking), followed by $n_{1d}$ canonical cluster updates. $n_{1d}$ and $n_{wolff}$  are set as described
in the previous section for the case of the pyrochlore lattice.

From the results for $A^{\chi}_{\psi_N}$ in Fig.~\ref{planpyrautos}, we see that the link percolation
based algorithm has very poor autocorrelations, and barely decorrelates configurations. The plaquette percolation based algorithm faces no
such difficulty. In the same figure, we also display results obtained for
the algorithm with dynamic premarking.
We see that the
that the dynamic premarking strategy described in Sec.~\ref{premarkingstrategies} leads to further, and quite significant improvements
in autocorrelations over the Swendsen-Wang type scheme that uses configuration-independent premarking.

\begin{figure}[t]
  \includegraphics[width=\columnwidth]{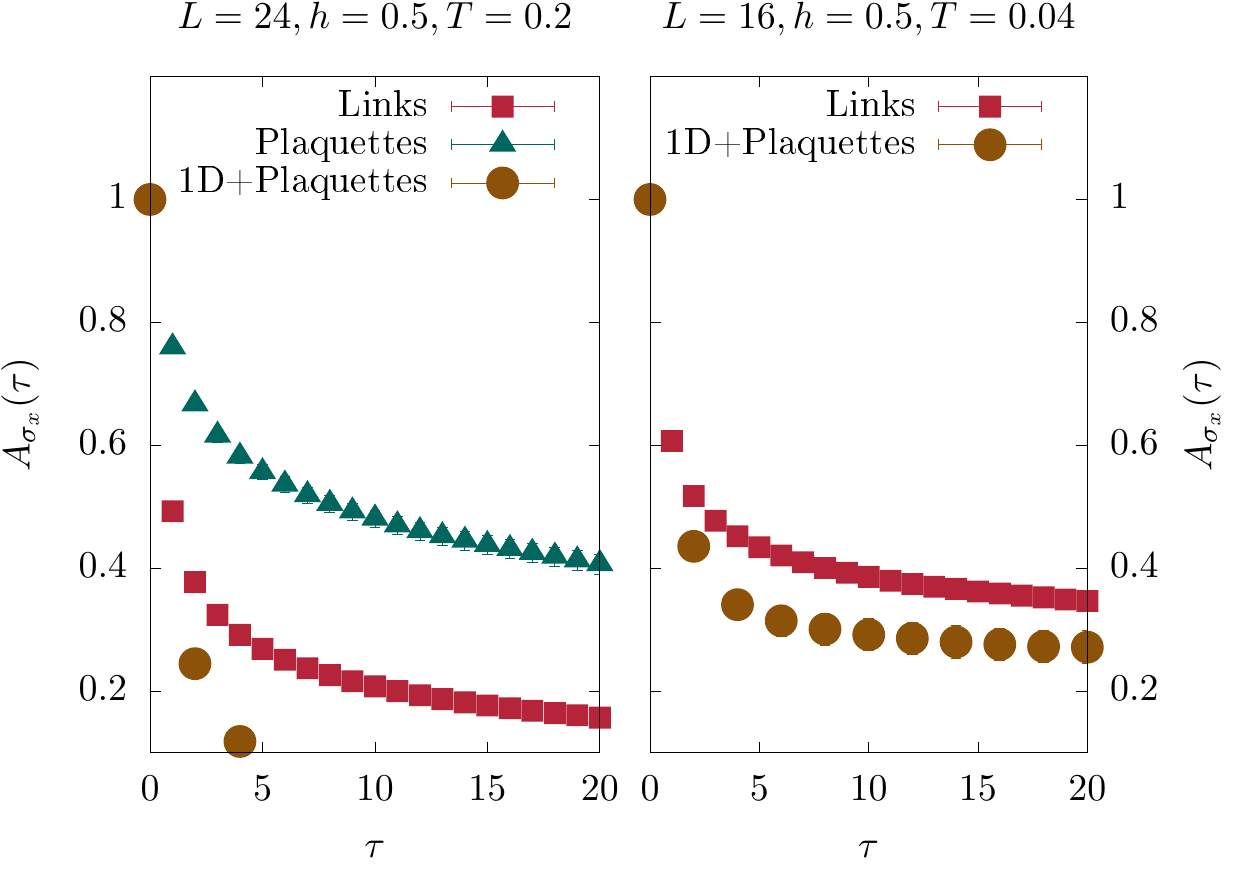}
  \caption{\label{sxplanpyrautos}Time series of Monte Carlo estimator of the equal time correlation function recorded during the simulation of the TFIM on a $L \times L$ planar pyrochlore lattice by the link percolation based cluster algorithm and the plaquette percolation based cluster algorithm with and without the canonical (`1D') updates.}
\end{figure}

We also measure the mean value of $\sigma^x$, and calculate the
corresponding autocorrelation function $A_{\sigma^x}$.
From the results displayed in Fig.~\ref{sxplanpyrautos}, we see that the conventional link
percolation based algorithm decorrelates the estimators of $\sigma^x$ somewhat better than the plaquette percolation based
algorithm if canonical updates are not used. However, when the plaquette percolation
based microcanonical clusters are supplemented with canonical updates,
this issue is resolved.
As shown in the same figure (Fig.~\ref{sxplanpyrautos}), such a combined approach
improves the autocorrelation times of $\sigma^x$ over those of the link percolation
based algorithm.
To make this comparison, we again scale the $x$-axis of $A_{\sigma^x}$ for the
combined algorithm by a 
factor of $2$ to account for the time required to perform the canonical updates in each Monte  Carlo step.

\begin{figure}[t]
  \includegraphics[width=\columnwidth]{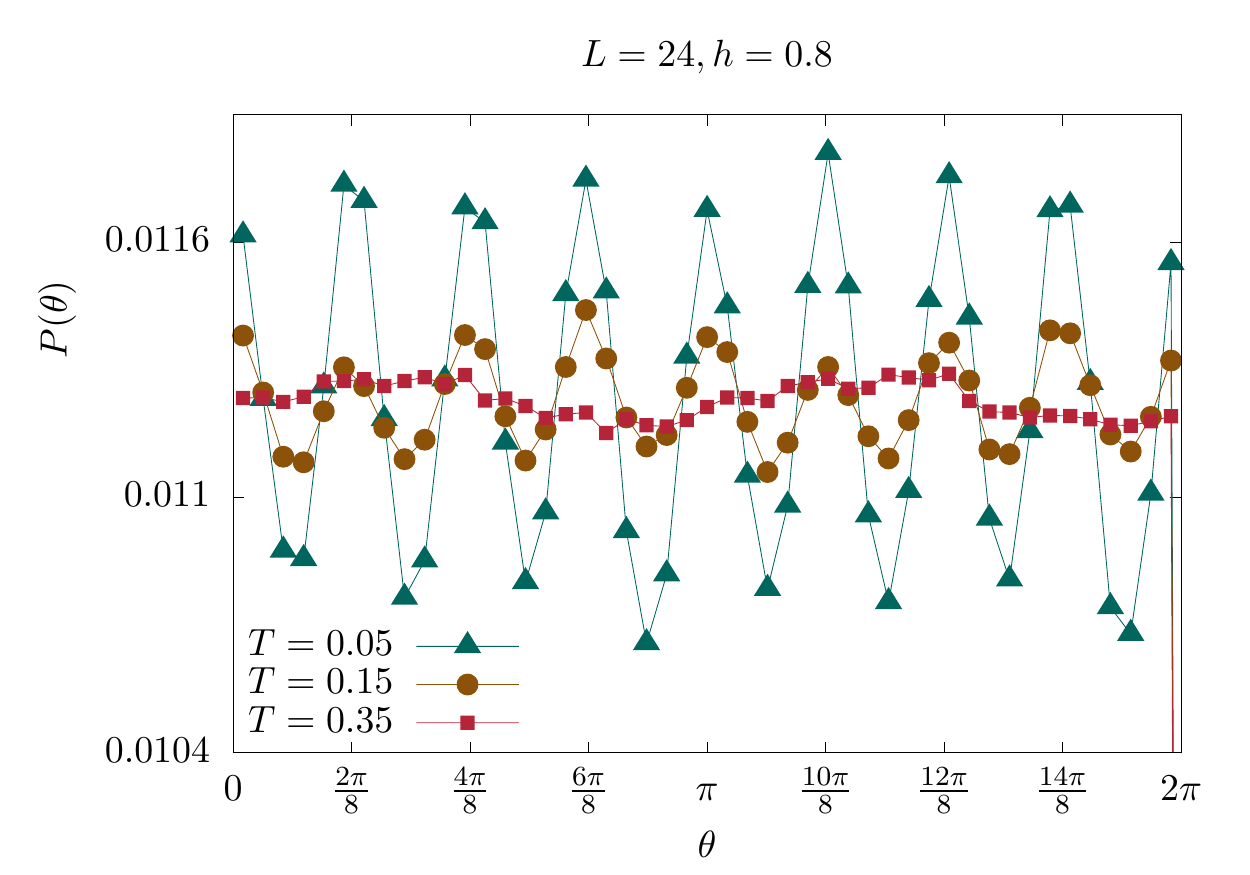}
  \caption{\label{melting}Histograms of the phase $\theta$ of the complex order parameter $\psi$ defined in Eq.~\eqref{fftfimoparam} for a FFTFIM on a $L \times L$ square lattice, obtained using the plaquette percolation based cluster algorithm. At low temperatures, eight peaks at
$\theta=2n\pi/8, n\in (0,7)$ characterize the eight-fold columnar order, while flat histograms of $\theta$ at higher temperatures characterize the emergent $U(1)$ symmetry.}
\end{figure}
\section{Illustrative Example}
\label{Results}
To demonstrate the possible uses of our algorithm, we look at the fully frustrated TFIM on the square lattice
.
This is known to host a columnar-ordered groundstate at not-too-large magnetic fields.\cite{Mila_fftfim} This groundstate order is characterized by the order parameter $\psi$,
defined in Eq.~\ref{fftfimoparam}. Corresponding to the eight columnar ordered states, $\psi$ takes the eight values
$\pm 1$, $\pm i$, $\pm(1 \pm i)$, up to an overall magnitude.
Following Blankschtein et al.,\cite{Blankschtein_fftfim} one can write down a Landau-Ginsburg functional of the order parameter $\psi$, containing the leading-order symmetry-allowed terms that control the physics:
\begin{equation}
  \mathcal{F}_{\mathrm{XY8}}(\psi)= |\psi|^2 +u_4|\psi|^4 +u_6|\psi|^6 + u_8 |\psi|^8 +v_8(\psi^8+\psi*^8).
  \label{lgaction_fftfim}
\end{equation}
On the basis of this Landau-Ginzburg theory, the finite temperature phase transitions of
the FFTFIM can be expected to lie in the universality class of the $XY$-model with an eight-fold symmetry breaking perturbation.\cite{Blankschtein_fftfim}
Such models were studied in Ref.~\onlinecite{Jose_etal} using renormalization group methods. From  such analyses, one expects
that the columnar order melts via a two-step melting process, with an intermediate temperature power-law ordered phase. 
We calculate the argument of the complex estimator of the order parameter $\psi$ in our Monte Carlo simulations and display its histograms in  Fig.~\ref{melting}. 
As is clear from this figure, the eight peaks corresponding to the eight ordered states  get washed away with increasing
temperature. The flat histograms at intermediate temperatures are indicative of the emergent $U(1)$ symmetry in the power-law ordered phase.

\begin{figure}[t]
  \includegraphics[width=\columnwidth]{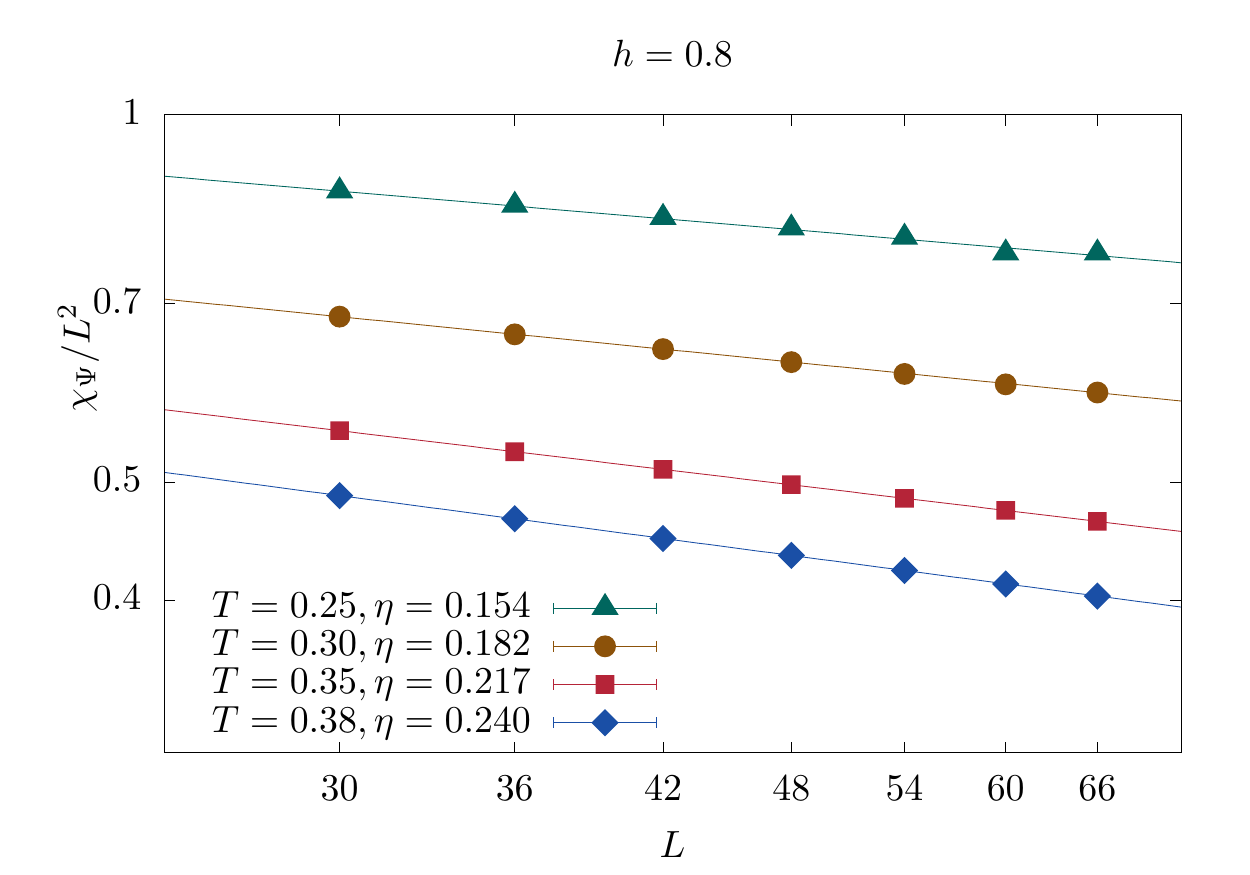}
  \caption{\label{etafits} Data for $\chi_{\psi}/L^2$  for the FFTFIM on a $L \times L$ square lattice, calculated in Monte Carlo simulations using the plaquette percolation based algorithm, fit well to the power-law form $kL^{-\eta}$ for a range of temperatures with the parameter $\eta$ within the expected range $(1/16,1/4)$, providing evidence for the intermediate power-law ordered phase associated with two-step melting.}
\end{figure}
To make this more precise, we we display power-law fits of $\chi_{\psi}/L^2$ for the FFTFIM on a  square lattice to 
the form $kL^{-\eta}$ for four different temperatures. From this analysis, we see that the best fit values of $\eta$ lie in the range
$(1/16,1/4)$ expected for such an intermediate power-law ordered phase.\cite{Jose_etal} This provides strong evidence for the intermediate power-law ordered
phase.

\section{Acknowledgements}
One of us (KD) gratefully acknowledges the organizers of The International Workshop on Low Dimensional Quantum Magnetism 2016 (EPFL Lausanne) and SIGN 2017 (Institute for Nuclear Theory, Seattle) for hospitality during the initial stages of this work, and Pranay Patil for a stimulating discussion on his ideas for more efficient canonical update procedures, beyond the simplest version discussed here. We gratefully acknowledge
the hospitality of ISSP Tokyo during the final stages of the work and the writing
of the initial draft of this manuscript. We also aknowledge extremely useful and stimulating discussions with S.~Wessel and P.~Emonts regarding the autocorrelation times of off-diagonal operators in the methods described here, particularly in comparison with the performance of their ``greedy'' variant\cite{Emont_Wessel} of our original triangular-plaquette based approach.\cite{Biswas_Rakala_Damle} Our computational work relied exclusively on the computational resources of the Department of Theoretical Physics at the Tata Institute of Fundamental Research (TIFR).

\appendix
\section{Distribution of cluster sizes.}
We presented a set of cluster-algorithms for frustrated Ising models which lead to significantly improved autocorrelation times. In the conventional algorithm, the clusters typically percolate and freeze, while the new set of algorithms produce a broader distribution of cluster sizes. This is the underlying cause of the improvement in performance. To illustrate this point, we calculate and compare the cluster size distributions produced by the new algorithms with premarking, and compare them with the ones produced by the conventional algorithms based on link-based decomposition.
As a measure of the size of the cluster, we consider the total number of vertex-legs (Sec.~\ref{review}) of the operator string in the cluster, $n_c$. For the Swendsen-Wang type updates performed in both the conventional algorithm based on link-based decomposition and the new algorithm based on plaquette-based decomposition with premarking, the cluster-size distribution $P(n_c)$ is estimated by $N_{n_c}/N_{\mathrm{tot.}}$, the number of clusters of size $n_c$ constructed by the algorithm in the course of the simulation, divided by the total number of clusters produced.  In the alternate Wolff-type construction (used when the premarking is done dynamically), the probability that a randomly chosen vertex-leg is part of a cluster of $n_c$ vertex-legs is proportional to $n_c$. The probability of making large clusters is thus enhanced by this factor in the Wolff updates when compared to Swendsen-Wang updates. Therefore, in this case, we define $P(n_c)$ to be proportional to $N_{n_c}/n_c$.

\begin{figure}[t]
  \includegraphics[width=8cm]{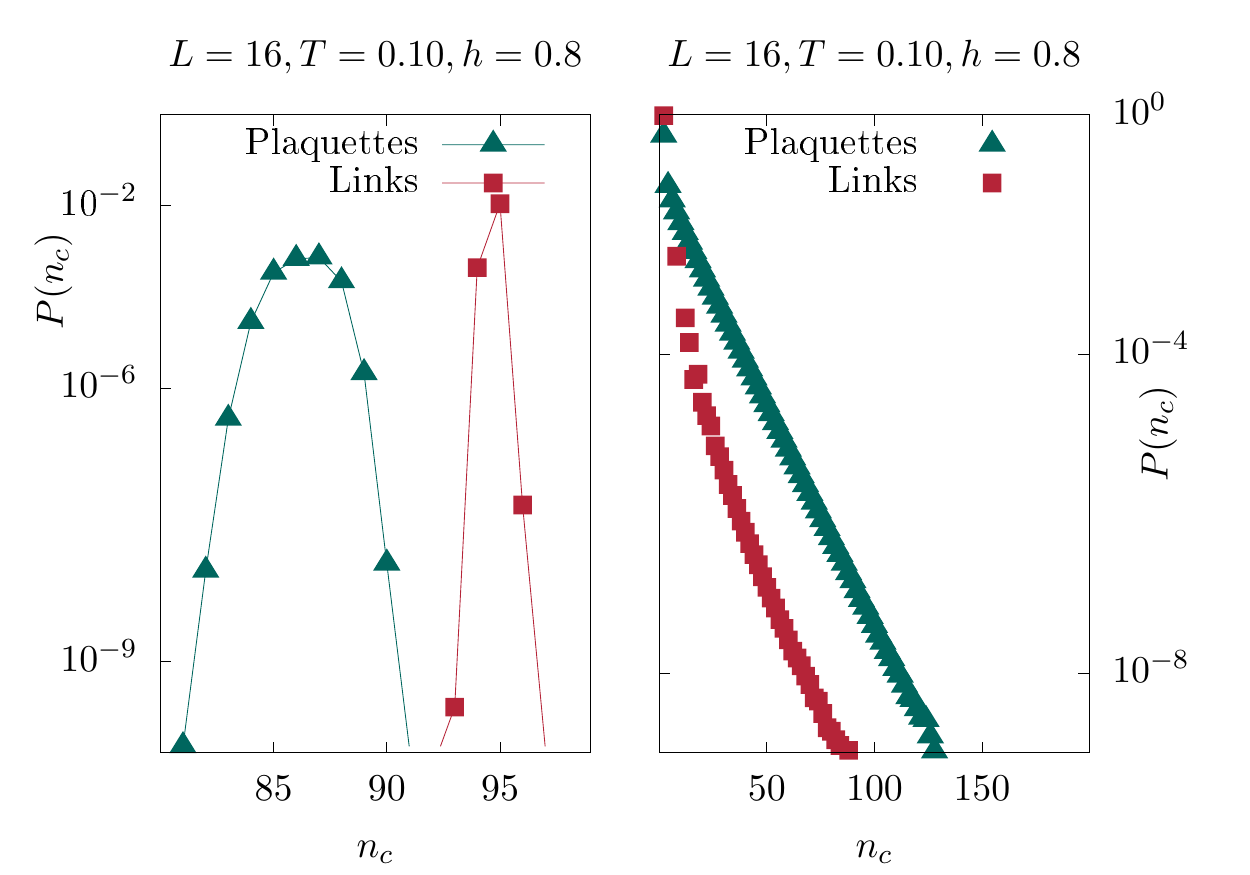}
  \caption{\label{fftfim_clusts}Distribution of cluster sizes in the FFTFIM on a $L\times L$ square lattice for large clusters (left panel) and small clusters (right panel) using both the link percolation based cluster algorithm and the
plaquette percolation based cluster algorithm.} 
\end{figure}

\begin{figure}[t]
  \includegraphics[width=8cm]{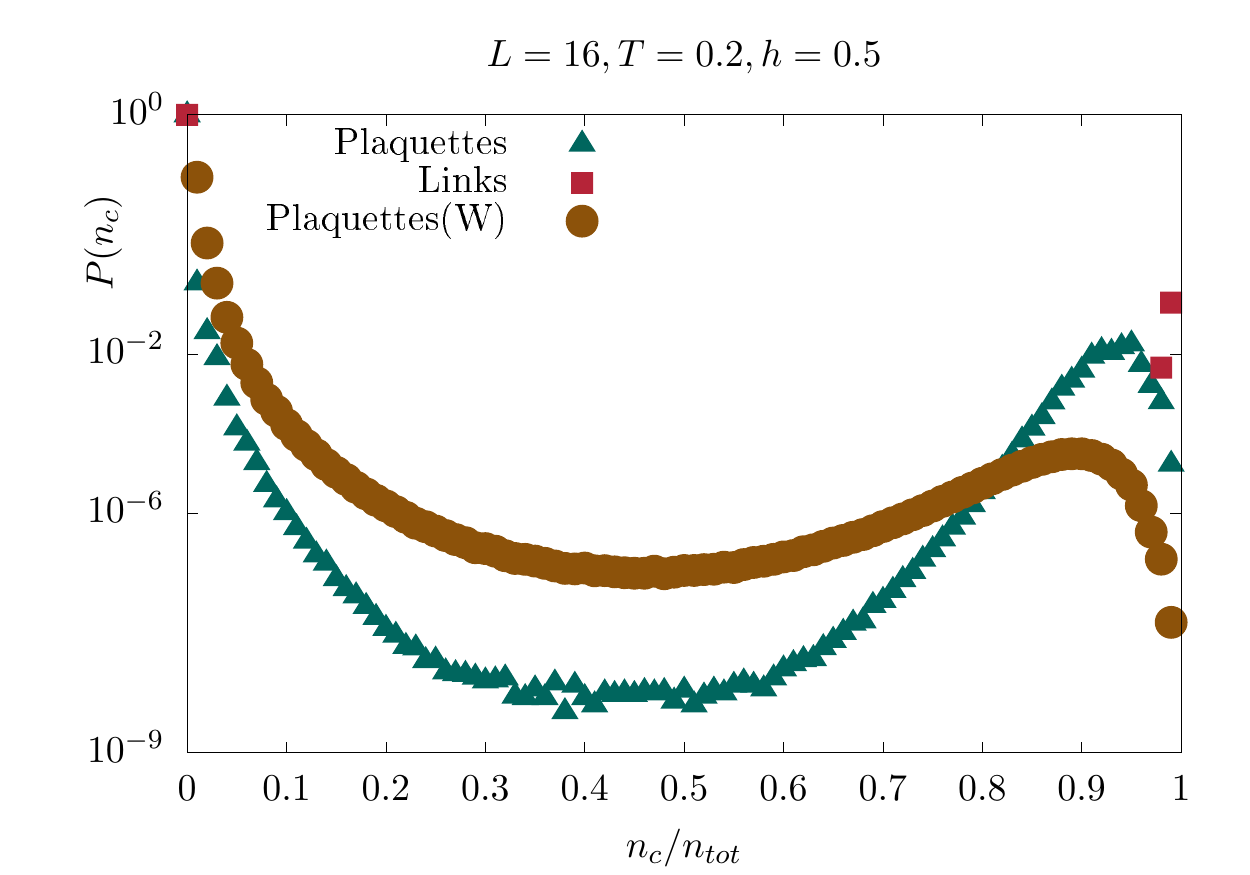}
  \caption{\label{planpyr_clusts} Distribution of cluster sizes for the TFIM on a $L \times L$ planar pyrochlore lattice for the link percolation based cluster algorithm and the plaquette percolation based cluster algorithm using both the Swendsen-Wang updates with configuration-independent premarking (labelled `Plaquettes') and the Wolff updates with dynamic premarking (labelled `Plaquettes(W)') described in Sec.~\ref{premarkingstrategies}.}
\end{figure}

\begin{figure}[t]
  \includegraphics[width=8cm]{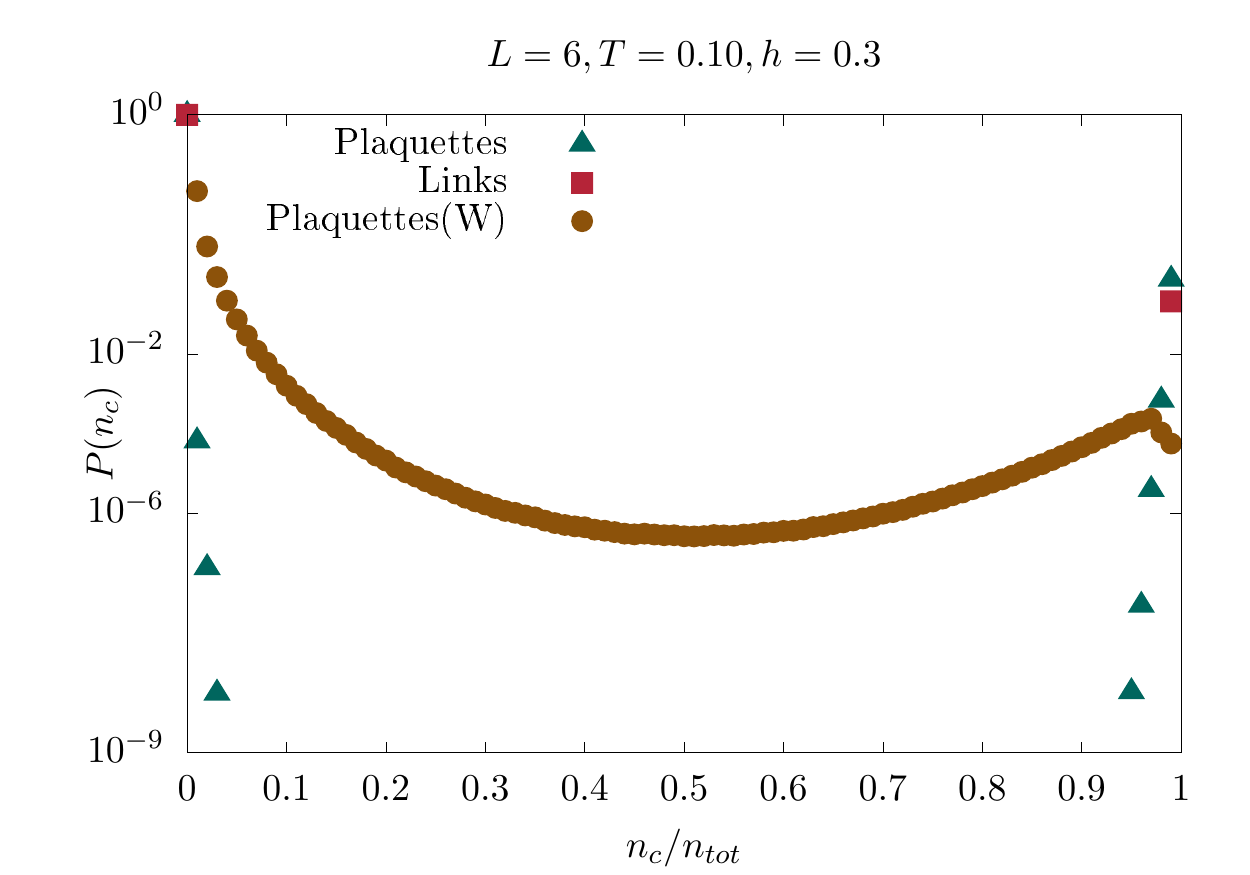}
  \caption{\label{pyr_clusts} Distribution of cluster sizes for the TFIM on a $L \times L \times L$ pyrochlore lattice for the link percolation based cluster algorithm and the plaquette percolation based cluster algorithm using both the Swendsen-Wang updates with configuration-independent premarking (labelled `Plaquettes') and the Wolff updates with dynamic premarking (labelled `Plaquettes(W)') described in Sec.~\ref{premarkingstrategies}.}
\end{figure}

We observe that the algorithm for the FFTFIM on the square lattice makes  makes two kinds of clusters, : \emph{small} and \emph{large},  reminiscent of the behaviour noted in Ref.~\onlinecite{Biswas_Rakala_Damle} for the TFIM on the triangular lattice. We have found the  distribution of small clusters to be quite robust to changes in system size and temperature for both the conventional and new algorithm. On the other hand, the large clusters seem to scale with the
total number of vertex-legs in the operator string, $n_{tot}$. We therefore show the distribution of large clusters as a function of $n_c/n_{tot}$. The cluster-size distributions are displayed in Fig.~\ref{fftfim_clusts}. We see that the distributions for both small and large clusters are considerably broader for the plaquette-based algorithm with premarking.
For the transverse field Ising models on the pyrochlore and planar pyrochlore lattices, the contrast between cluster size distributions produced by the conventional algorithm and the new algorithm with premarking is more dramatic. The cluster-size distribution $P(n_c)$ obtained using the link-based algorithm have non-zero values essentially at very small and very large values of $n_c$, consistent with their inability to effeciently decorrelate configurations, as established in Sec.~\ref{Performance}. The plaquette-based algorithm with premarking has a significantly broader distribution of cluster sizes, while the Wolff-updates with dynamic premarking makes further, more significant improvements. The cluster-size distributions for the TFIM on the planar pyrochlore lattice are displayed in Fig.~\ref{planpyr_clusts}, while the cluster-size distributions for the TFIM on the pyrochlore lattice are shown in Fig~\ref{pyr_clusts}. The data-points for the link algorithm correspond to isolated points at both ends of the range of the $x$-axis.

\section{The importance of premarking.}
\label{imp_premarking}
\begin{figure}[t]
  \includegraphics[width=\columnwidth]{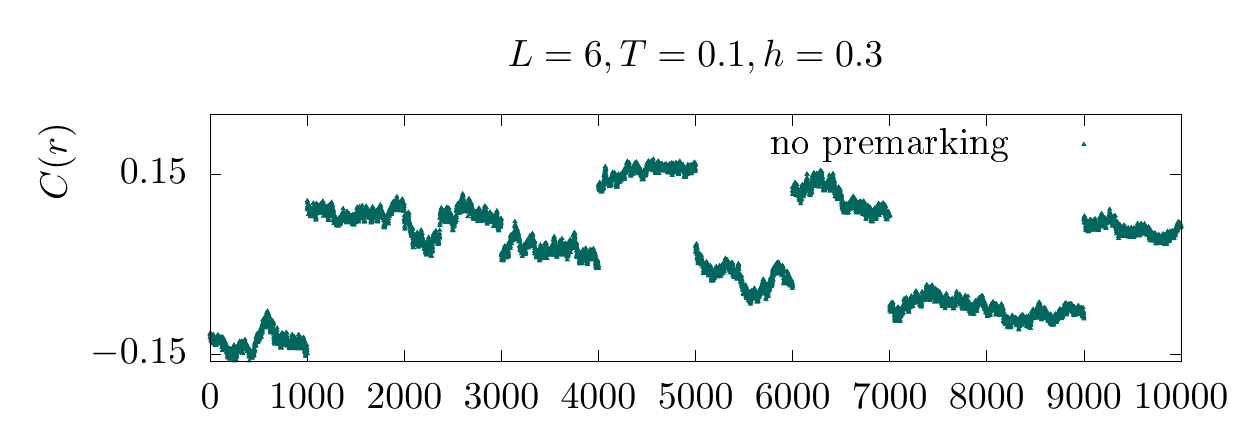}
  \caption{\label{sztimes2}Time series of Monte Carlo estimator of $C$, the equal time correlation function of $\sigma^z$, recorded during the simulation of the TFIM on a $L \times L \times L$ pyrochlore lattice, obtained by a plaquette percolation based cluster algorihtm without premarking, as outlined in Sec.~\ref{imp_premarking}}
\end{figure}
\begin{figure}[t]
  \includegraphics[width=\columnwidth]{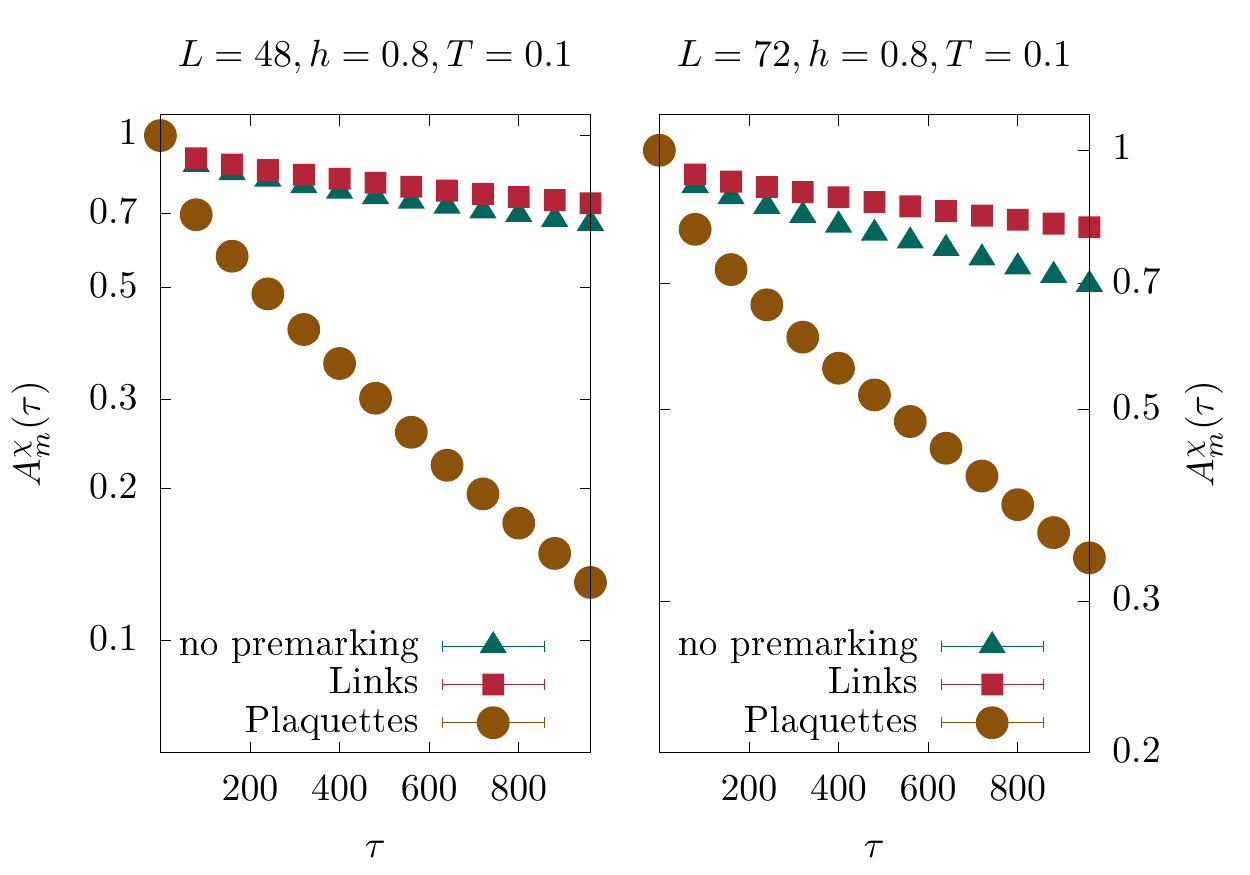}
  \caption{\label{triangular}Comparison of autocorrelation function for the Monte Carlo estimator of $\chi_{m}$ (where $m=\sum_{i}\sigma^{z}_i$ is the uniform magnetization)  in the TFIM on a $L \times L$ triangular lattice using both the link percolation based cluster algorithm and plaquette percolation based cluster algorithms with and without premarking.}
\end{figure}
We have mentioned in Sec.~\ref{needmicrocan} that the efficiency of the microcanonical cluster updates presented in this paper derives from two ingredients : \emph{plaquette-decomposition} of the Ising exchange part of the Hamiltonian(into triangles,squares or tetrahedra), and \emph{premarking motifs on the lattice} to ensure consistency of the clusters over imaginary times and hence broader cluster-size distributions.

To illustrate the crucial role played by the premarked motifs, we implement a cluster algorithm for the TFIM on the pyrochlore lattice without premarking. We decompose the Ising exchange part of the Hamiltonian into terms living on tetrahedra as in Eq.~\eqref{pyr_decomp}. However, we now decompose the Ising vertices of weight $6J$ according to any one (randomly chosen) decompositions chosen from the three which separate  two antiparallel spins into a cluster, and the other two parralel spins into another cluster. Similarly, for the Ising exchange vertices of weight $8J$, we randomly choose one of the two possible microcanonical decompositions which separate the vertex into two clusters, each with a  pair of antialigned spins. We display the time series of Monte Carlo estimators of the equal-time correlation function $C$, defined in Eq.~\eqref{corrfnsz}, obtained using this algorithm in Fig.~\ref{sztimes2}. Comparison with Fig.~\ref{sztimes} shows that this algorithm suffers from loss of ergodicity as severe as the conventional algorithm with link-based decomposition. Thus the premarked motifs introduced here are key.

Similarly, we have implemented the algorithm for TFIM on the triangular lattice presented in Ref.~\onlinecite{Biswas_Rakala_Damle} without premarking. This is equivalent to the greedy variant of the approach of Ref.~\onlinecite{Biswas_Rakala_Damle}, which was tested recently\cite{Emont_Wessel} and found to have a performance not significantly different from the standard link-based approach. We have displayed autocorrelations of the uniform susceptibility using this algorithm and compared it to the algorithm with premarking in Fig.~\ref{triangular}; these results again underscore the importance with premarking.

\section{Dominant and subdominant order parameters for the fully frustrated TFIM on the square lattice}
In this section, we motivate the order parameter $\psi$ appearing in the Landau-Ginzburg-Wilson action of the FFTFIM on the square lattice (Eq.~\ref{lgaction_fftfim}).
\subsection{Soft-spin analysis.}
\label{softmodes}
First, we perform a soft-spin analysis to identify the lowest energy modes of the classical Ising-exchange part of the Hamiltonian defined in Eq.~\ref{fftfim_eqn}, repeating the analysis of Blankschtein et al.~\cite{Blankschtein_fftfim} for the specific \emph{gauge} choice in our Hamiltonian. Relaxing the normalization contstraint ($\sigma_i^2=1$) on the spins and Fourier transforming the Hamiltonian, we get
\begin{align}
  \frac{1}{L^2}\sum_{\vec{k}}\mathbf{s}(\vec{k})^{\dagger}J(\vec{k})\mathbf{s}(\vec{k}) ,\\
  \mathbf{s}(\vec{k})=(s_{I}(\vec{k}),s_{II}(\vec{k})),\\
  J(\vec{k})=\frac{1}{2}J\left(
  \begin{array}{cc}
    2\cos(k_y)  & -(1+e^{-ik_x}) \\
    -(1+e^{+ik_x})  &-2\cos(k_y)  \\
  \end{array}
  \right) \; .
\end{align}
The subscripts $I$ and $II$ refer to the two basis sites in each unit cell of the Hamiltonian.
The eigenvalues of $J(\vec{k})$ are given by $\pm (J/2)\sqrt{4\cos^ 2(k_y)+2(1+cos(k_x))}$. The minimum eigenvalues are $-J\sqrt{2}$, and correspond to the  
two degenerate modes $\vec{k}=(0,0)$ and $k=(0,\pi)$. The corresponding eigenvectors are $(\sin(\pi/8),\cos(\pi/8))$ and $(\cos(\pi/8),\sin(\pi/8))$ respectively.
We construct two order parameters corresponding to these modes:
\begin{align}
  \psi^{(0,0)}=\sigma_{I}(0,0)\sin(\pi/8)+\sigma_{II}(0,0)\cos(\pi/8) \\
  \psi^{(0,\pi)}=\sigma_{I}(0,\pi)\cos(\pi/8)+\sigma_{II}(0,\pi)\sin(\pi/8)
  \label{fftfimoparam2}
\end{align}
Using the spin densities $\sigma_A$, $\sigma_B$, $\sigma_C$ and $\sigma_D$ corresponding to the four-sublattice decomposition of the square lattice  instead of 
$\sigma_{I}(0,0)$ and $\sigma_{II}(0,\pi)$, and combinbing the two order parameters in Eq.~\ref{fftfimoparam2} into a single complex order parameter as 
$\psi=\psi^{(0,0)}+i\psi^{(0,\pi)}$ gives us Eq.~\ref{fftfimoparam}.

\subsection{Analysis of representation carried by the sublattice spin densities.}

Although this standard soft-spin analysis certainly identifies the dominant order parameter correctly, we find it instructive to supplement it with a symmetry analysis that is more natural from the computational point of view. The starting point of this alternate approach is the computational result that the low temperature state corresponds to a four sublattice decomposition of the square lattice, and an ordered state is characterized by the values of $\sigma_{\alpha}$ with $\alpha=A,B,C,D$. Given this, it is natural to ask what representation of the symmetry group of the microscopic Hamiltonian  is carried by
the four dimensional vector $\sigma_\alpha$, and decompose it into irreducible representations. This is outlined in this appendix.

In general, the problem of determining 
the irreps appearing in the LGW theory given the low-symmetry group $H$ (preserved by the ordered state) and the high-symmetry group $G$ of symmetries of the microscopic Hamiltonian (such that $H\in G$) is called the `Inverse Landau Problem'.\cite{ascher1,ascher2}. 
We need a much less general analysis.
From our QMC simulations we find that the low-temperature phase and of the FFTFIM on the square lattice  corresponds to the spin densities $\sigma_{I}(0,0)$, $\sigma_{I}(0,\pi)$, $\sigma_{II}(0,0)$ and $\sigma_{II}(0,\pi)$ acquiring non-zero values, where the subscripts denote the the basis sites of the two-site unit cell of the Hamiltonian. This is consistent with earlier QMC studies of the ground state,~\cite{Mila_fftfim} the cartoon groundstates of Sec~\ref{models}, and the soft-spin analysis of the previous subsection. We know that these four spin-densities carry a four-dimensional representation of the group $G$, $\Upsilon$. It is obvious that the order parameter appearing in the LGW theory corresponding to this transition must be constructed out of these four spin-densities. So, we decompose this representation $\Upsilon$ into irreps of $G$. For our purposes, it suffices to consider the auxillary group $G/Ker(\Upsilon)\simeq Im(\Upsilon)$, which is the group formed by the matrices in the image of $\Upsilon$. The representation spaces carrying the irreps of $Im(\Upsilon)$ are the same spaces that carry the irreps of the group $G$.
To determine the matrices of the representation $\Upsilon$, we look at the action of the group $G$ on the four spin-densities. For convenience we reorganize the basis of four spin-densities into $\sigma_A, \sigma_B, \sigma_C, \sigma_D$, where the subscripts refer to the four-sublattice decomposition of the square lattice (Fig.~\ref{fftfim_lat}).
 We present a generating set for the group $Im(\Upsilon)$ defined by the representation 
$\Upsilon$ : the identity $I$, global spin flip $I_{-1}$, translation by one lattice spacing along $\hat{y}$ $T_y$,  translation along $\hat{x}$ along with simultaneous flip 
of relevant spins $T^{G}_x$, and rotation about the origin by $\pi/2$ along with simulatenous flip of relevant spins $R^{G}_{\pi/2}$.
The actions of these elements on the spin-densities on the four sublattices are given by 
\begin{align}
\nonumber I_{-1} : (\sigma_A, \sigma_B, \sigma_C, \sigma_D) &\rightarrow (-\sigma_A, -\sigma_B, -\sigma_C, -\sigma_D)\\
\nonumber T_y : (\sigma_A, \sigma_B, \sigma_C, \sigma_D) &\rightarrow (\sigma_D, \sigma_C, \sigma_B, \sigma_A) \\
\nonumber T^{G}_x : (\sigma_A, \sigma_B, \sigma_C, \sigma_D) &\rightarrow (-\sigma_B, -\sigma_A, \sigma_D, \sigma_C) \\
R^{G}_{\pi/2} : (\sigma_A, \sigma_B, \sigma_C, \sigma_D) &\rightarrow (-\sigma_A, \sigma_D, \sigma_C, \sigma_B). 
\label{generatingset}
\end{align}
Note that group elements $g$ which belong to the low-symmetry subgroup $H$ ( $g\in H$) are represented by the identity matrix.
We calculate the character table of $Im(\Upsilon)$ using Burnside's method,~\cite{burnside,dixon} which we now proceed to describe
briefly. We start with the lemma that for a conjugacy class $R$ and an irrep $\rho$, one has $\sum_{r\in R} \rho(r)=|R|\chi(r)/d$, where 
$\chi(r)$ is the character for the conjugacy class $r$ and $d$ is the dimension of the representation.
To prove this, one can use Schur's lemma with the fact that $\sum_{r\in R}\rho(r)$ commutes with all matrices $\rho(g)$, for $g\in G$. Considering two different conjugacy classes $R$ and $S$, we have
\begin{align}
\nonumber  \sum_{r\in R}\rho(r) \sum_{s \in S} \rho(s) =\sum_{r\in R,s\in S} \rho(rs)\\
\implies \frac{|R|\chi(r)}{d}\frac{|S|\chi(s)}{d}=\sum_{T}c_{RST}\frac{\chi(t)|T|}{d},
\label{blemma}
\end{align}
where $c_{RST}$ is the number of pairs with $r\in R, s\in S$ for for a specific $t \in T$ (the number is independent of the choice of $t$).
Consider the matrices $X^{i}$, $X^{i}_{jk}=c_{R_i R_j R_k}$. It is clear form Eq.~\ref{blemma}, that the matrix $X^{i}_{jk}$  has eigenvalues 
$\chi_{r_i}|R_i|/d$, for any $r_i\in R_i$. The corresponding eigenvector has components $v_j=\chi(r_j)|R_j|/d$. Therefore, we can calculate
the characters of the all irreps by calculating simulatenous eigenvectors of the matrices $X^{i}$.

Using the character table obtained, we find that $Im(\Upsilon)$ has three two-dimensional representations, and three one-dimensional representations. Our four dimensional representation $\Upsilon$ decomposes into two
distinct two-dimensional representations. Both of these two dimensional representations are faithful, and are therefore candidates for the irrep controlling 
the behaviour of the transition(s) to a broken-symmetry state. To determine the order parameters, we  construct the projectors into the subspaces carrying these irreps. We find that the first projector
projector projects to the real and imaginary parts of the complex order parameter $\psi$  defined in Eq.~\ref{fftfimoparam}.
The second projects to another two-dimensional subspace spanned by a different, secondary complex order parameter $\phi$, given by
\begin{align}
\nonumber  \phi =& \frac{1}{L^2} [\sigma_A^{z}\exp(i \pi/8) +\sigma_B^{z}\exp(-i 5\pi/8) \\
  &+\sigma_C^{z}\exp(i 5\pi/8) +\sigma_D^{z}\exp(-i\pi/8) ].
  \label{fftfimoparam3}
\end{align}
As an aside, we note that $Im(\Upsilon)\simeq D_{16}$, the Dihedral group of order $16$. $D_{16}$ has three two-dimensional
irreps, usually called $A1$, $A2$, $B1$ and $B2$, and four one dimensional irreps $E1$, $E2$ and $E3$. Decomposing $\Upsilon$, we find
$\Upsilon= E_1 \oplus E_3$. Both $E_1$ and $E_3$ are faithful irreps of $D_{16}$.
$E_1$ corresponds to the order parameter $\psi$, while $E_3$ corresponds to the secondary order parameter $\phi$.

Under the action of the generating set of the group $Im(\Upsilon)$ presented in Eq.~\eqref{generatingset}, the order parameters $\psi$ and $\phi$
transform as follows :
\begin{align} 
\nonumber I_{-1} : \psi\rightarrow -\psi; \phi\rightarrow -\phi \\
\nonumber T_y : \psi\rightarrow \psi^{*}; \phi\rightarrow\phi^{*}\\
\nonumber T^{G}_x: \psi\rightarrow e^{-i\pi/2} \psi^{*}; \phi\rightarrow e^{i\pi/2}\phi^{*}\\
R^{G}_{\pi/2} :  \psi\rightarrow e^{-i \pi/4} \psi^{*}; \phi\rightarrow e^{-i3\pi/4}\phi^{*}\\
\end{align}

The Landau-Ginzburg theory obtained by extending Eq.~\eqref{lgaction_fftfim} to include terms invariant in both $\psi$ and $\phi$ is:
\begin{align}
\nonumber  \mathcal{F}(\psi,\phi) = \mathcal{F}_{XY8}(\psi) + \mathcal{G}_{XY8}(\phi) +\mathcal{F}_{c}(\psi, \phi) \\
\nonumber  \mathcal{F}_{c}(\psi, \phi)= c_{22} |\psi|^2 |\phi|^2 + c_{13} (\psi \phi^{*3} +\psi^* \phi^{3}) \\
+c_{31} (\phi \psi^{*3} +\phi^* \psi^{3})
\label{extendedlgw}
\end{align}
$\mathcal{F}_{XY8}$ is the Landau-Ginzburg action of the XY model with an eightfold symmetry breaking perturbation introduced in Eq.~\eqref{lgaction_fftfim}, while $\mathcal{G}_{XY8}$ has the same form, with different values for the coefficients.
The transition in question corresponds to the ordering of $\psi$, modeled by a change in the character of the minimum of $\mathcal{F}_{XY8}(\psi)$. 
We note that the secondary order parameter in our gauge is expected to play the role of the primary order parameter in a formulation that uses a different gauge, in which the Hamilontian is given by Eq.~\ref{fftfim_eqn}, but 
$J_{ij}=|J|$ on the straight lines and $J_{ij}=-|J|$ on the zig-zag lines in Fig.~\ref{fftfim_lat}. In this other gauge, the Landau-Ginzburg theory can be obtained from
the Landau-Ginzburg theory of Eq.~\eqref{extendedlgw} by the switching the coupling constants appearing in $\mathcal{G}_{XY8}(\phi)$  with those of $\mathcal{F}_{XY8}(\psi)$,
 and by switching $c_{13}$ with $c_{31}$.

\begin{figure}[t]
  \includegraphics[width=8cm]{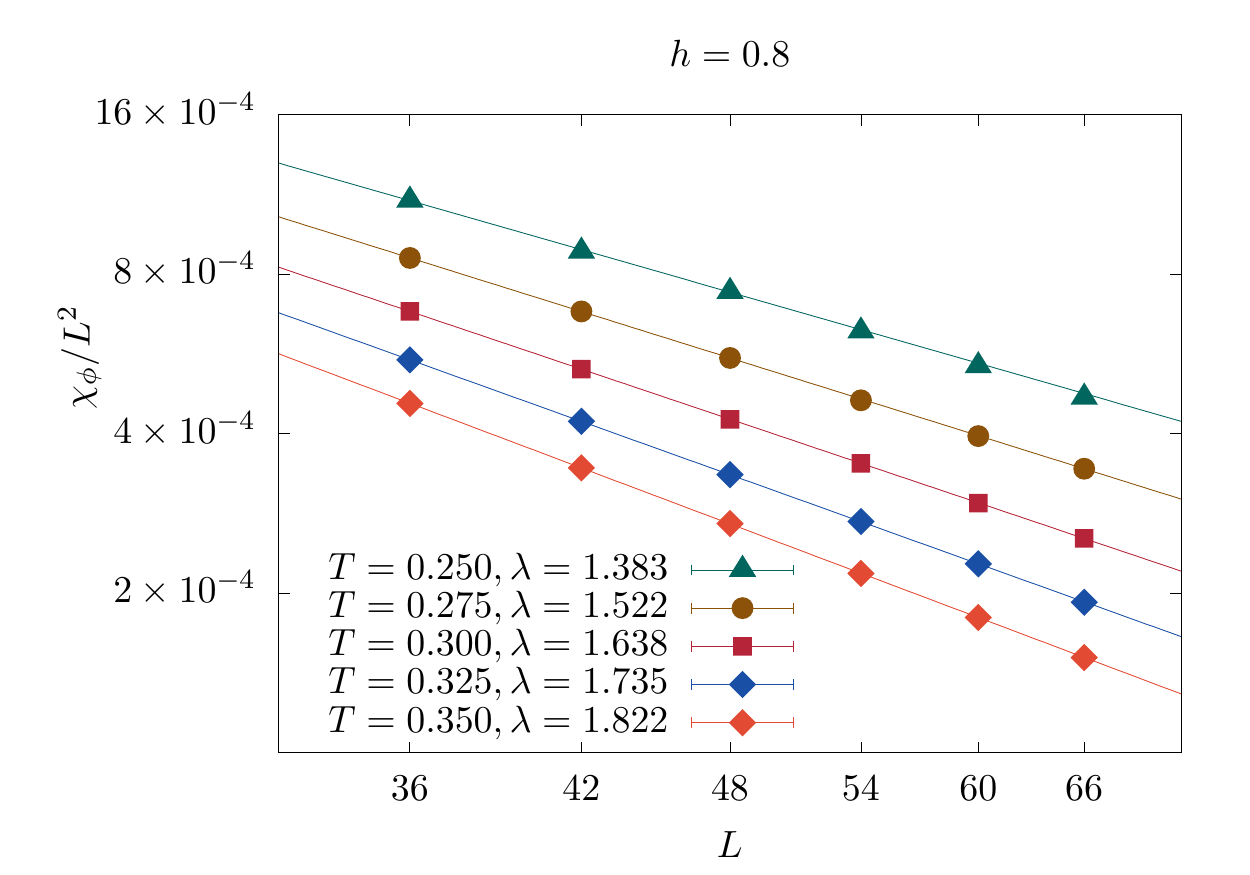}
  \caption{\label{etafits2} Data for $\chi_{\phi}/L^2$, the suceptibility corresponding to the secondary order parameter $\phi$ for the FFTFIM on a $L \times L$ square lattice, calculated in Monte Carlo simulations using the plaquette percolation based algorithm, fits well to the power-law form $kL^{-\lambda}$ in the intermediate power-law ordered phase associated with two-step melting.}
\end{figure}

\begin{figure}[t]
  \includegraphics[width=8cm]{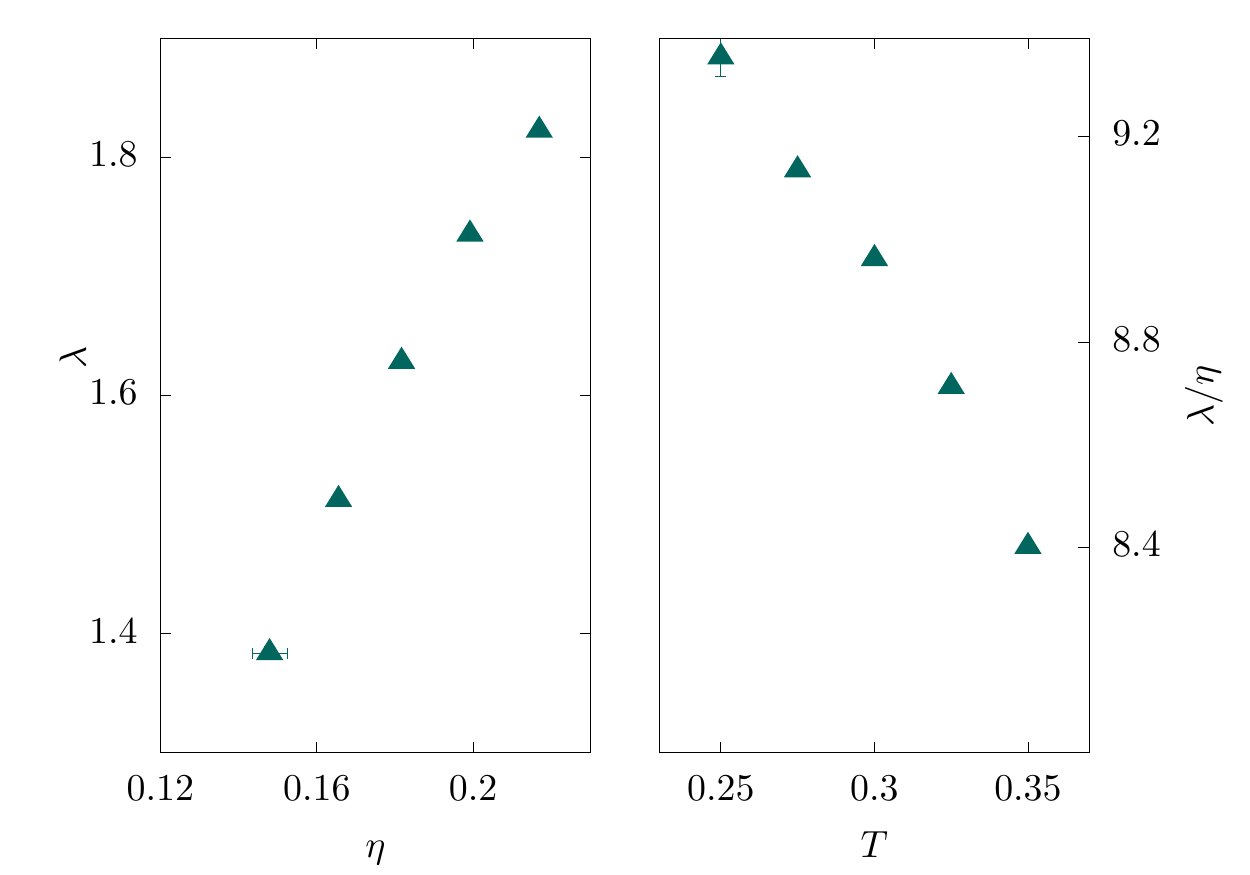}
  \caption{\label{lambdaplot} $\lambda$, the exponent corresponding to the secondary order parameter $\phi$, depends continuously on the exponent $\eta$ corresponding to the
primary order parameter $\psi$ (left panel). The temperature dependence of the ratio $\lambda/\eta$ in the intermediate power-law ordered phase (right panel).} 
\end{figure}

 Given that the Landau-Ginzburg description allows for such a coupling between $\phi$
 and $\psi$, it is clearly of interest to ask if correlations of $\phi$ reflect the criticality of $\psi$ in the two-step melting region.
 To study this, we measure $\chi_{\phi}/L^{2}$ and find that it seems to fit well to the power-law form $k' L^{-\lambda}$ in the intermediate power-law ordered phase where $\chi_{\psi}/L^2$ fits to $k L^{-\eta}$. The fits are displayed in Fig.~\ref{etafits2}. We have also displayed the exponent $\lambda$ corresponding to the secondary order-parameter $\phi$ as a function of $\eta$ in Fig.~\ref{lambdaplot}. 
 It would be interesting to understand this relationship between the two exponents on the basis of the Landau-Ginzburg theory of Eq.~\eqref{extendedlgw}, and we hope to return to this in future work.
\end{document}